%% file: main.tex
\pdfoutput=1
\documentclass[aps,10pt,prd,twocolumn,floats,letterpaper,showpacs,nofootinbib,bibnotes,notitlepage,superscriptaddress,floatfix]{revtex4-1}

\usepackage{graphicx}
\usepackage{mathrsfs}
\usepackage[intlimits,centertags]{amsmath}
\usepackage{amssymb,amsfonts}
\usepackage[pdftex]{hyperref}
\usepackage[x11names]{xcolor}
\usepackage{enumerate}
\usepackage{xfrac}
\usepackage[normalem]{ulem}
\usepackage{float}

\hypersetup{pdftitle={IceCube Data for Neutrino Point-Source Searches: Years 2008-2018},
pdfsubject={IceCube Data for Neutrino Point-Source Searches: Years 2008-2018},
pdfauthor={IceCube Collaboration},
pdfstartview={FitH},
colorlinks=true,
bookmarksopen=false,
bookmarksnumbered=false,
bookmarksopenlevel=0,
linkcolor=Blue1!60!black,
citecolor=Green1!50!black,
urlcolor=Blue1!70!black}

\newcommand{\MA}[1]{{\color{black}#1}}

\begin{document}

\title{IceCube Data for Neutrino Point-Source Searches: Years 2008--2018}

\input{authorlist.tex}
\collaboration{IceCube Collaboration$^{\dagger\dagger}$}
{\let\thefootnote\relax\footnote{{$^{\dagger\dagger}$Email: \href{mailto:analysis@icecube.wisc.edu}{analysis@icecube.wisc.edu}}}}
\noaffiliation


\begin{abstract}
IceCube has performed several all-sky searches for point-like neutrino sources using track-like events, including a recent time-integrated analysis using 10 years of IceCube data. This paper accompanies the public data release of these neutrino candidates detected by IceCube between April 6, 2008 and July 8, 2018. The selection includes through-going tracks, primarily due to muon neutrino candidates, that reach the detector from all directions, as well as neutrino track events that start within the instrumented volume. An updated selection and reconstruction for data taken after April 2012 slightly improves the sensitivity of the sample. While more than 80\% of the sample overlaps between the old and new versions, differing events can lead to changes relative to the previous 7 year event selection. An a posteriori estimate of the significance of the 2014-2015 TXS flare is reported with an explanation of observed discrepancies with previous results. This public data release, which includes 10 years of data and binned detector response functions for muon neutrino signal events, shows improved sensitivity in generic time-integrated point source analyses and should be preferred over previous releases.
\end{abstract}

\maketitle

\section{Introduction}

The IceCube Observatory~\cite{Aartsen:2016nxy} at the geographic South Pole has been operating at full capacity for the past ten years. In 2013, IceCube reported first evidence of an isotropic flux of astrophysical neutrinos in the TeV-PeV energy range~\cite{Aartsen:2013bka,Aartsen:2013jdh}. While the flux is by now observed with high significance~\cite{Aartsen:2014gkd,Aartsen:2015rwa,Aartsen:2016xlq,Aartsen:2020csc} its astrophysical origin remains uncertain~\cite{Ahlers:2018fkn}. In parallel, IceCube has been searching for high-energy neutrino emission from individual time-integrated point sources, including unbiased all-sky searches~\cite{Abbasi:2010rd,Aartsen:2013uuv,Aartsen:2014cva,Aartsen:2014cva,Aartsen:2018ywr,Aartsen:2019fau} as well as individual source candidates like cores of active galaxies~\cite{Bradascio:2019xdd}, blazars~\cite{Aartsen:2016lir}, $\gamma$-ray bursts~\cite{Abbasi:2012zw,Aartsen:2014aqy,Aartsen:2016qcr,Aartsen:2017wea}, fast radio bursts~\cite{Aartsen:2017zvw}, nearby galaxies~\cite{Aartsen:2019xkn}, diffuse Galactic emission~\cite{Albert:2018vxw}, Galactic $\gamma$-ray sources~\cite{Kheirandish:2019bke}, pulsar wind nebulae~\cite{Aartsen:2020eof}, or X-ray binaries~\cite{Abbasi:2011ke}. IceCube also takes part in various realtime multi-messenger activities via its fast response to external alerts in photons~\cite{Abbasi:2011ja,Aartsen:2015trq} or gravitational waves~\cite{Adrian-Martinez:2016xgn,ANTARES:2017iky,Aartsen:2020mla} and by reporting astrophysical neutrino candidate events~\cite{Aartsen:2016lmt,AyalaSolares:2020ing}. Recently, IceCube was able to report first compelling evidence of neutrino emission from the $\gamma$-ray blazar TXS 0506+056~\cite{Finley:2019vpk,IceCube:2018dnn,IceCube:2018cha}.

In order to encourage engagement with the broader multi-messenger astrophysics community, IceCube is dedicated to releasing datasets for public use. Data has been released in various formats \cite{IC40data, IceCube:2013hese, IC59data, IceCube:2015numu, IceCube:2015hese4, IceCube:2016ps1, IceCube:2018, IceCube:TXS2018, IceCube:Alerts2018, IceCube:2019, IceCube:2019pst}. To better equip the community, IceCube is releasing a new 10 year dataset of track-like events previously developed for IceCube's recent 10~year time integrated point source search \cite{Aartsen:2019fau} along with binned instrument response functions to describe the detector. 

This paper accompanies the public data release of track-like neutrino candidates detected by IceCube between April 6, 2008 and July 8, 2018 \cite{data:IC40IC86VII}. The released sample shows evidence of cumulative excess of events from a catalogue of 110 sources with respect to the expected atmospheric neutrino background. Its significance of 3.3$\sigma$ is mostly determined by four sources, in order of importance NGC 1068, TXS 0506+056, PKS 1424+240 and GB6 J1542+6129. NGC~1068, a Seyfert galaxy at a redshift of z=0.003, is spatially coincident with the hottest spot in the full Northern sky search.

The underlying event selection, called {\tt PSTracks} in the following, is designed for point-source studies that benefit from the good angular resolution of tracks and can tolerate larger atmospheric background contributions compared to diffuse neutrino analyses. The {\tt PSTracks} sample has recently been updated from 7~years ({\tt v2}) to 10~years ({\tt v3}). {\tt PSTracks} v3, which includes an improved selection and reconstruction for data collected after April 2012, is now being released to the community.

This data release applies to IceCube data collected prior to July 2018 and may be used to reproduce analyses published on the {\tt PSTracks} v3 dataset. The IceCube Collaboration continues to evaluate and refine the selection, reconstructions, and calibrations for internal use and regular future public releases including these updates will be provided.

In the following, we give an account of the IceCube detector (Sec.~\ref{secI}), the event selection (Sec.~\ref{secII}), and detector performance (Sec.~\ref{secIII}). We also highlight changes to previous data selections and releases (Sec.~\ref{secIV}).

\section{The IceCube Observatory}\label{secI}

The IceCube Observatory identifies neutrino interactions in the vicinity of the detector by the Cherenkov light emitted by relativistic charged secondary particles traveling through the deep ultra-clear glacial ice. The in-ice detector consists of 5,160 Digital Optical Modules (DOMs) that are distributed across a cubic kilometer of glacial ice at the South Pole~\cite{Abbasi:2008aa, Abbasi:2010vc}. The DOMs are distributed on 86 read-out and support cables (``strings'') and are deployed between 1.45~km and 2.45~km below the surface. Most strings follow a triangular grid with a width of 125~m, evenly spaced over the volume. The DOMs consist of a photomultiplier tube, electronics for digitization, and LEDs for detector calibration~\cite{Abbasi:2008aa, Abbasi:2010vc}.

The main IceCube array has a neutrino energy threshold of about 100~GeV. Contained within the main IceCube detector, a denser array in the clearest glacial ice, known as DeepCore, lowers the energy threshold to about 10~GeV.
The IceCube Observatory also includes a surface array of 82 pairs of water Cherenkov detectors, called IceTop~\cite{IceCube:2012nn}, that detects and reconstructs cosmic ray air showers above $300\:$TeV. 

Neutrino interactions observed in the IceCube array generally may either follow a track-like or cascade-like topology. The \emph{track}-like signal events originate primarily in charged-current interactions of muon (anti-)neutrinos ($\nu_\mu$ \& $\bar{\nu}_\mu$) with nucleons producing energetic muons. Tau (anti-)neutrinos ($\nu_\tau$ \& $\bar{\nu}_\tau$) may also produce charged-current interactions leading to energetic muons. Below $700$~GeV, secondary muons lose energy mainly due to ionization; above $700$~GeV, stochastic energy losses due to radiative emission become the dominant component. At TeV energies, muons travel long distances, larger than several kilometers in the Antarctic ice~\cite{Chirkin:2004hz}. Light is constantly emitted along the track. The resulting long lever arm allows precise directional reconstruction with median angular resolution $\Delta\Psi<1^\circ$. The absolute pointing accuracy of IceCube has been demonstrated to be $\lesssim0.2^\circ$~\cite{Aartsen:2013zka} via measurements of the effect of the Moon shadow on the background cosmic ray (CR) flux. 

Charged-current interactions of astrophysical electron or tau (anti-)neutrinos, as well as neutral current interactions of any neutrino type, primarily produce \emph{cascade}-like signal events with an almost spherically symmetric Cherenkov light emission, resulting in a median angular resolution of $\sim10^\circ$--$15^\circ$\cite{Aartsen:2017eiu}. These signal event signatures are, in general, less useful for point-source studies, except in searches for soft neutrino sources in the Southern hemisphere~\cite{Aartsen:2019epb}, extended Galactic neutrino emission~\cite{Aartsen:2019epb}, and transient sources, such as $\gamma$-ray bursts~\cite{Aartsen:2016qcr}. Moreover, the number of track-like events greatly outnumbers that of cascade-like events because neutrinos can interact far outside the detector prior to the detection of the secondary muon with IceCube.  

\begin{table}[t]
\centering
\begin{ruledtabular}
\begin{tabular}{lllccc}
\multicolumn{6}{c}{Data Samples} \\[0.1cm]
Season & Start & End & Livetime & Events & Ref.\\ 
IC40 & 2008/04/06 & 2009/05/20 & 376.4\,d&  36900 & \cite{Abbasi:2010rd} \\
IC59 & 2009/05/20 & 2010/05/31 & 352.6\,d& 107011 &  \cite{Aartsen:2013uuv}\\
IC79 & 2010/06/01 & 2011/05/13 & 316.0\,d&  93133 & \cite{Schatto:2014kbj}\\
IC86-I & 2011/05/13 & 2012/05/15 & 332.9\,d & 136244 & \cite{Aartsen:2014cva} \\
IC86-II & 2012/04/26\footnotemark & 2013/05/02 & 332.0\,d & 112858 & \cite{Aartsen:2019fau} \\
IC86-III & 2013/05/02 & 2014/05/06 & 362.9\,d & 122541 &  \cite{Aartsen:2019fau}\\
IC86-IV & 2014/04/10\footnotemark & 2015/05/18 & 370.7\,d& 127045 & \cite{Aartsen:2019fau}\\
IC86-V & 2015/04/24\footnotemark & 2016/05/20 & 365.4\,d & 129311 & \cite{Aartsen:2019fau}\\
IC86-VI & 2016/05/20 & 2017/05/18 & 357.3\,d& 123657 & \cite{Aartsen:2019fau}\\
IC86-VII & 2017/05/18 & 2018/07/08 & 410.6\,d& 145750 & \cite{Aartsen:2019fau}\\
\end{tabular}
\footnotetext[1]{Start of test runs of new processing; remainder of this season began 2012/05/15}
\footnotetext[2]{Start of test runs of new processing; remainder of this season began 2014/05/06}
\footnotetext[3]{Start of test runs of new processing; remainder of this season began 2015/05/18}
\end{ruledtabular}
\caption[]{IceCube seasons with corresponding start and end dates, lifetime, and total event numbers. We also indicate references in which the sample selection is described in detail.}\label{tab:livetimes}
\end{table}

The majority of the background of the {\tt PSTracks} sample originates from CRs interacting with the atmosphere to produce showers of particles including atmospheric muons and neutrinos. The atmospheric muons from the Southern hemisphere are able to penetrate the ice and are detected as track-like events in IceCube at a rate orders of magnitude higher than the corresponding atmospheric neutrinos~\cite{Aartsen:2016nxy}. Atmospheric neutrinos also produce muons from charged-current muon (anti-)neutrinos interactions, acting as an irreducible background in both hemispheres. Atmospheric muons from the Northern hemisphere are filtered out by the Earth. 

This data release includes events that were observed during the last three years of the construction phase of the IceCube observatory. These data seasons are referred to as IC40, IC59, and IC79 in Table~\ref{tab:livetimes}, with the names corresponding to the number of installed detector strings. Years following detector completion are referred to as IC86 with a numeral indicating the years since completion. Selections, software, and calibrations used by IceCube varied through these years until being standardized for seasons starting in IC86-II. 

{\tt PSTracks} v3 includes updates to the selection and reconstruction for IC86-II through IC86-IV along with three additional years of data. More than 80\% of the events from these seasons pass both v2 and v3 selections. The IC40~\cite{IC40data}, IC59~\cite{IC59data}, IC79, and IC86-I\cite{data:IC40IC86VII} selections remain unchanged relative to the {\tt PSTracks} v2 sample and are identical to previously released versions of each dataset. These years are included in the current data release in order to standardize formatting and to provide effective area and descriptions of detector responses.

Below, we briefly describe the event selection used in the released samples. See Tab.~\ref{tab:livetimes} references for further details about each sample. Additionally, we describe the response of the {\tt PSTracks v3} selection to signal of through-going muons from cosmic neutrinos and to atmospheric backgrounds. For more detailed information, refer to \cite{Abbasi:2010rd}, \cite{Aartsen:2013uuv}, and \cite{Aartsen:2014cva}.



\begin{figure}[t]
\centering
\includegraphics[width=\linewidth,viewport= 0 0 530 390, clip=true]{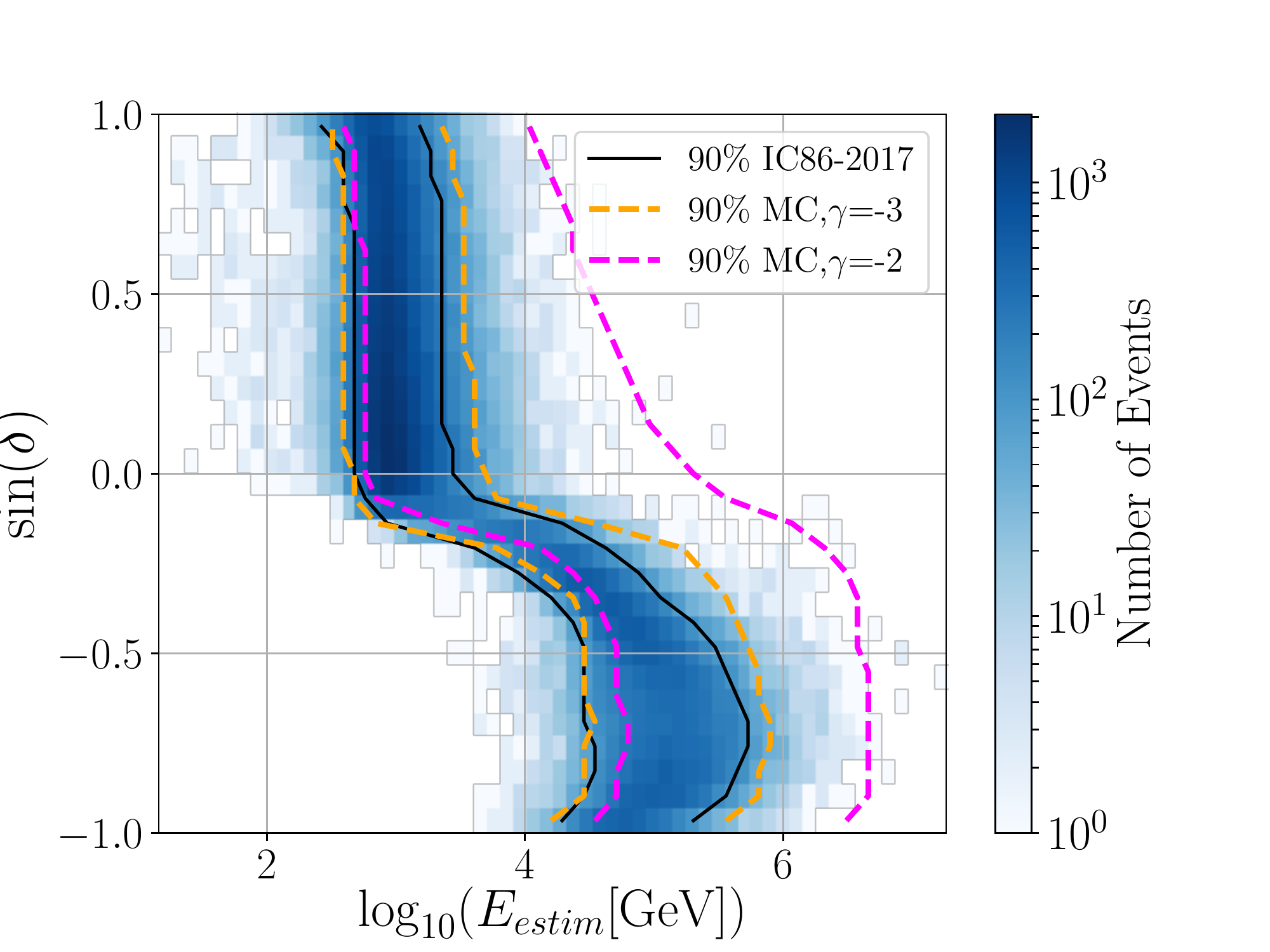}
\caption[]{The distribution of events in one year of data for the final event selection as a function of reconstructed declination and estimated energy. The 90\% energy range for the data (black), as well as simulated astrophysical signal Monte-Carlo (MC) for an $E^{-2}$ and an $E^{-3}$ spectrum are shown in magenta and orange respectively as a guide for the relevant energy range of IceCube (from Ref.~\cite{Aartsen:2019fau}).}\label{fig:enDist}
\end{figure}

\section{Event Selection}\label{secII}

The {\tt PSTracks} event selection identifies high-energy muons passing through the IceCube detector with a goal of identifying sources of astrophysical neutrinos. The selection applies differing criteria in the Northern and Southern hemisphere -- corresponding to events originating below (``up-going'') and above (``down-going'') the IceCube detector, respectively -- with different atmospheric backgrounds. The boundary between the hemispheres is at declination $\delta=-5^\circ$, which is identical to a zenith angle of $95^\circ$ for the special location of IceCube. 

In the Northern hemisphere, atmospheric muons are filtered by the Earth. While some atmospheric muons are erroneously reconstructed into the Northern sky, the misreconstructed events can be removed by selecting high-quality track-like events. 

In the Southern hemisphere, the atmospheric background is reduced by strict cuts on the reconstruction quality and minimum energy, since the astrophysical neutrino fluxes are expected to have a harder energy spectrum than the background of atmospheric muons and neutrinos. 

\begin{figure*}[t]\centering
\begin{minipage}[c][6.5cm][c]{0.49\textwidth}\centering
\includegraphics[width=\linewidth]{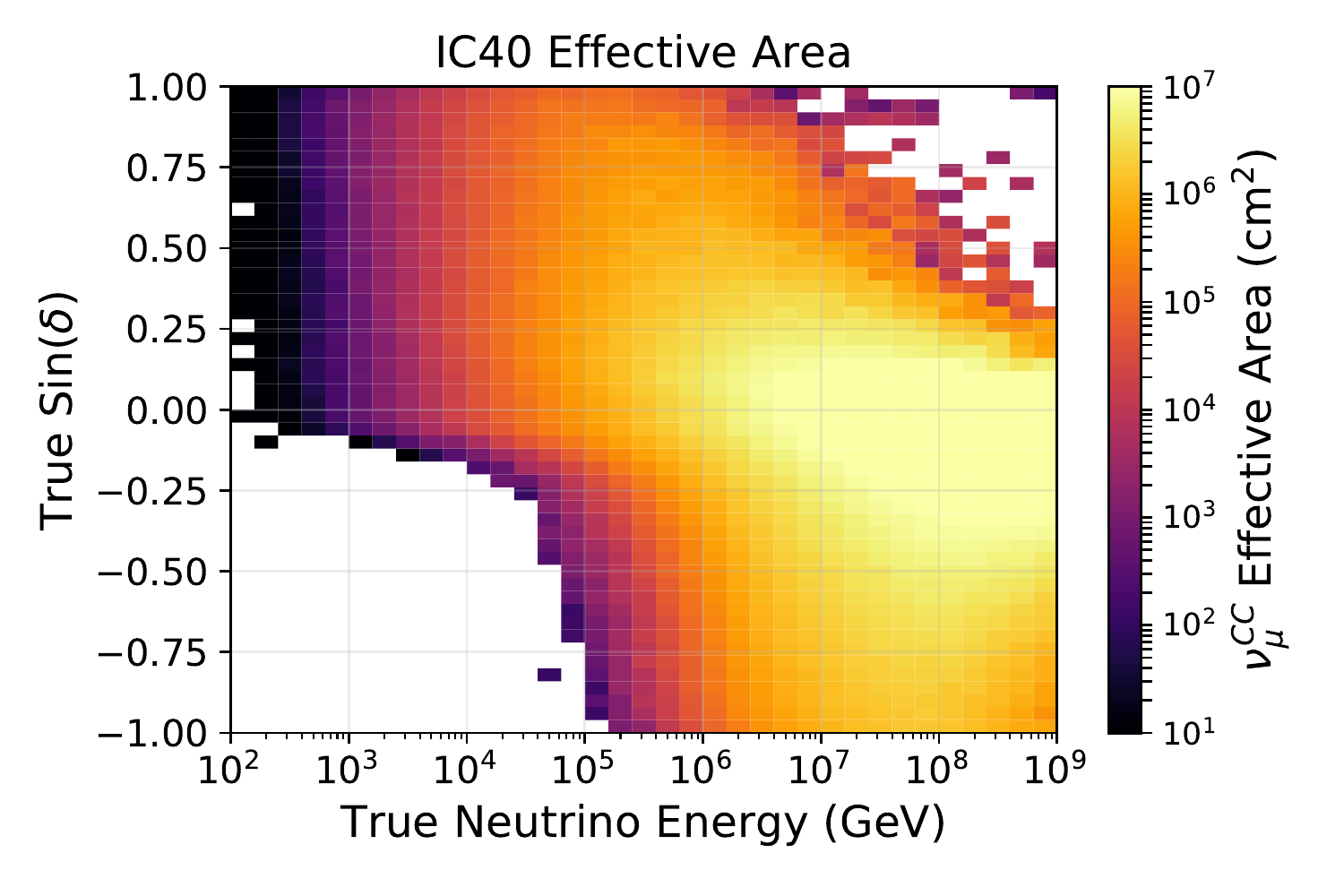}
\end{minipage}
\begin{minipage}[c][6.5cm][c]{0.49\textwidth}\centering
\includegraphics[width=\linewidth]{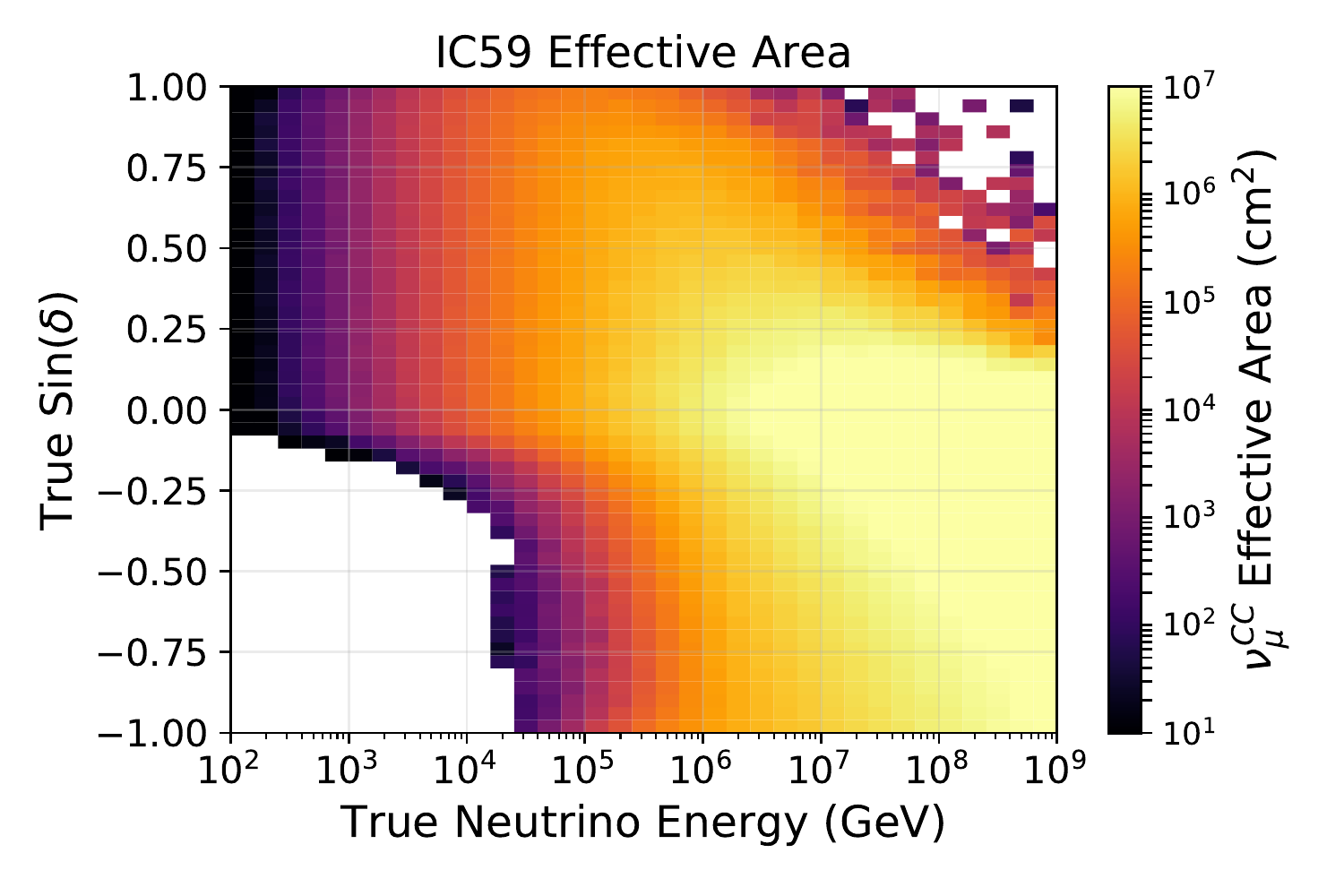}
\end{minipage}\\
\begin{minipage}[c][6.5cm][c]{0.49\textwidth}\centering
\includegraphics[width=\linewidth]{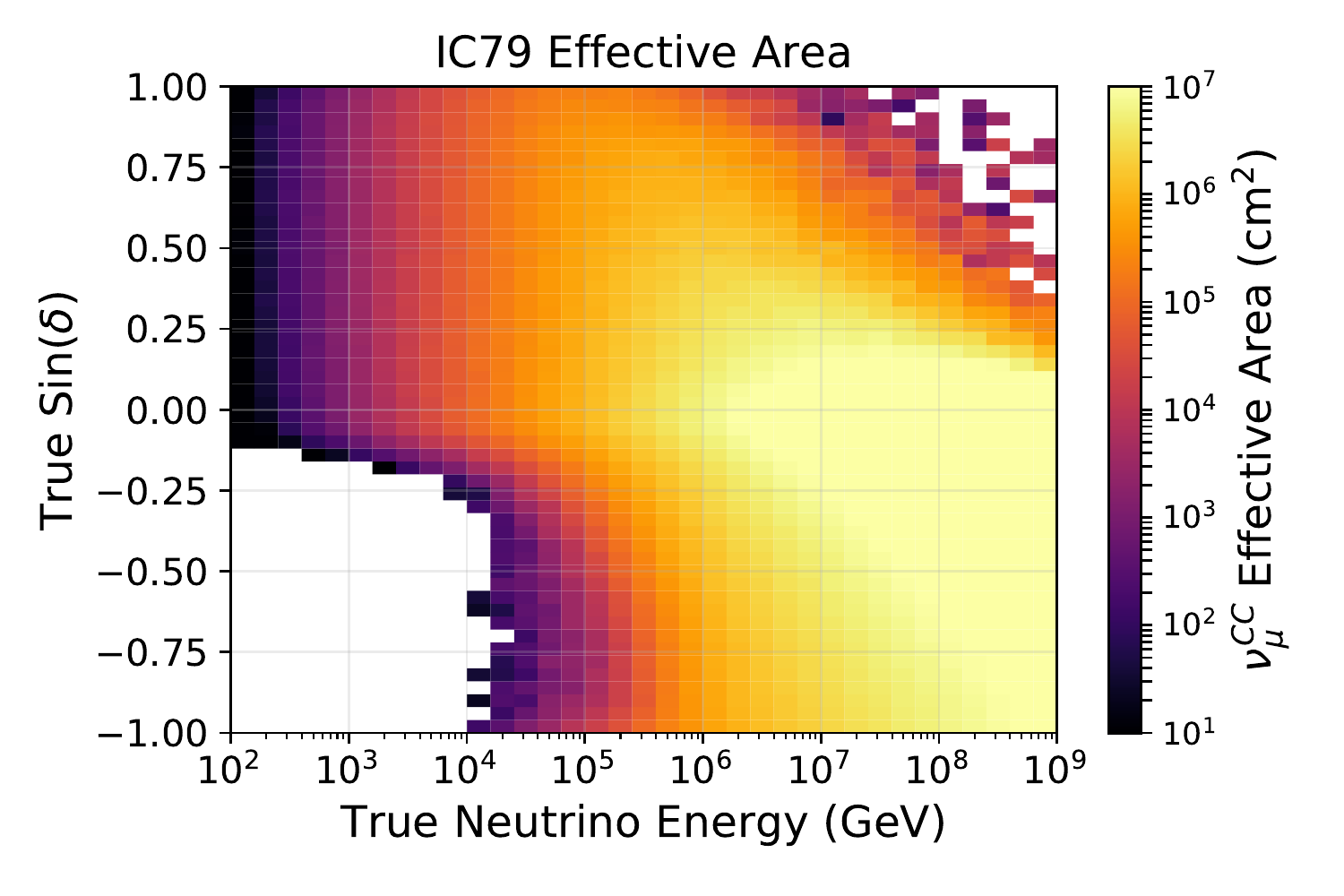}
\end{minipage}
\begin{minipage}[c][6.5cm][c]{0.49\textwidth}\centering
\includegraphics[width=\linewidth]{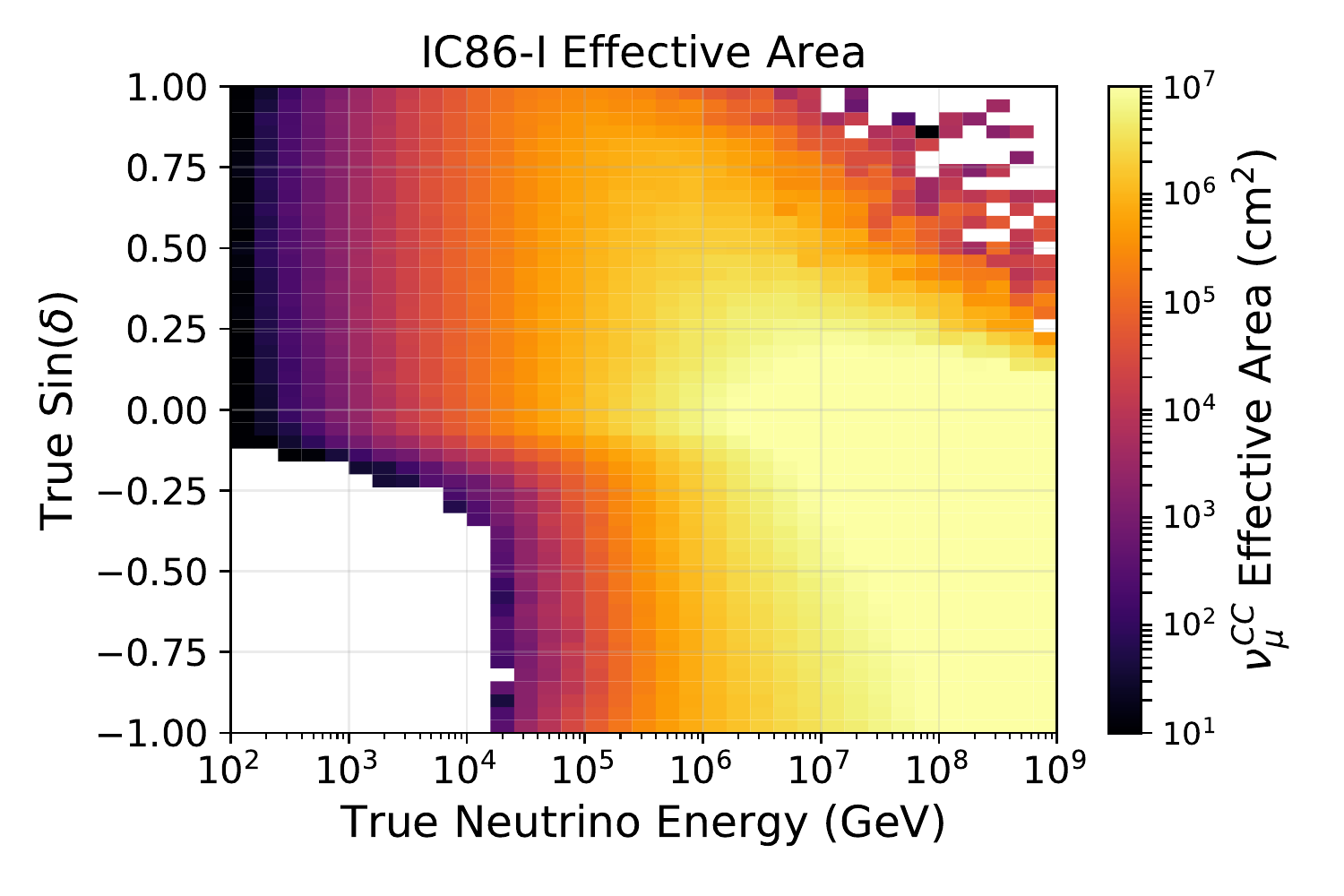}
\end{minipage}\\
\begin{minipage}[c][6.5cm][c]{0.49\textwidth}\centering
\includegraphics[width=\linewidth]{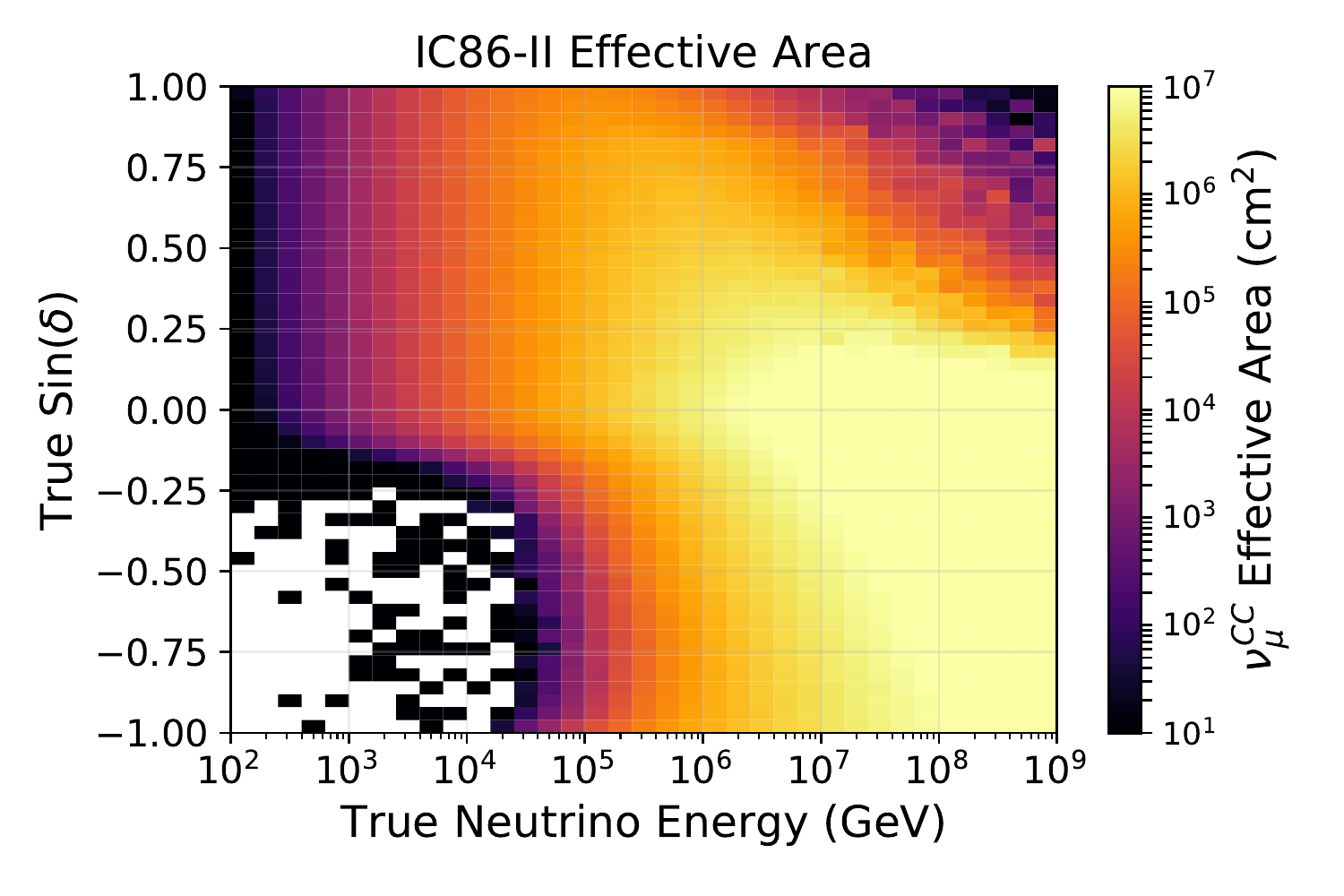}
\end{minipage}

\caption{The effective areas for each uniform data-taking period included in the current data release. IC40 through IC79 used the partially-completed detector while IC86-I and later followed detector completion. Software and calibration differences separate IC86-I from later years. Differences in the tools and applied selection have a small, but important impact on the sample for each period.}\label{fig:effA}
\end{figure*}

\begin{figure*}[t]\centering
\begin{minipage}[c][3.5cm][c]{0.32\textwidth}\centering
\includegraphics[width=\linewidth]{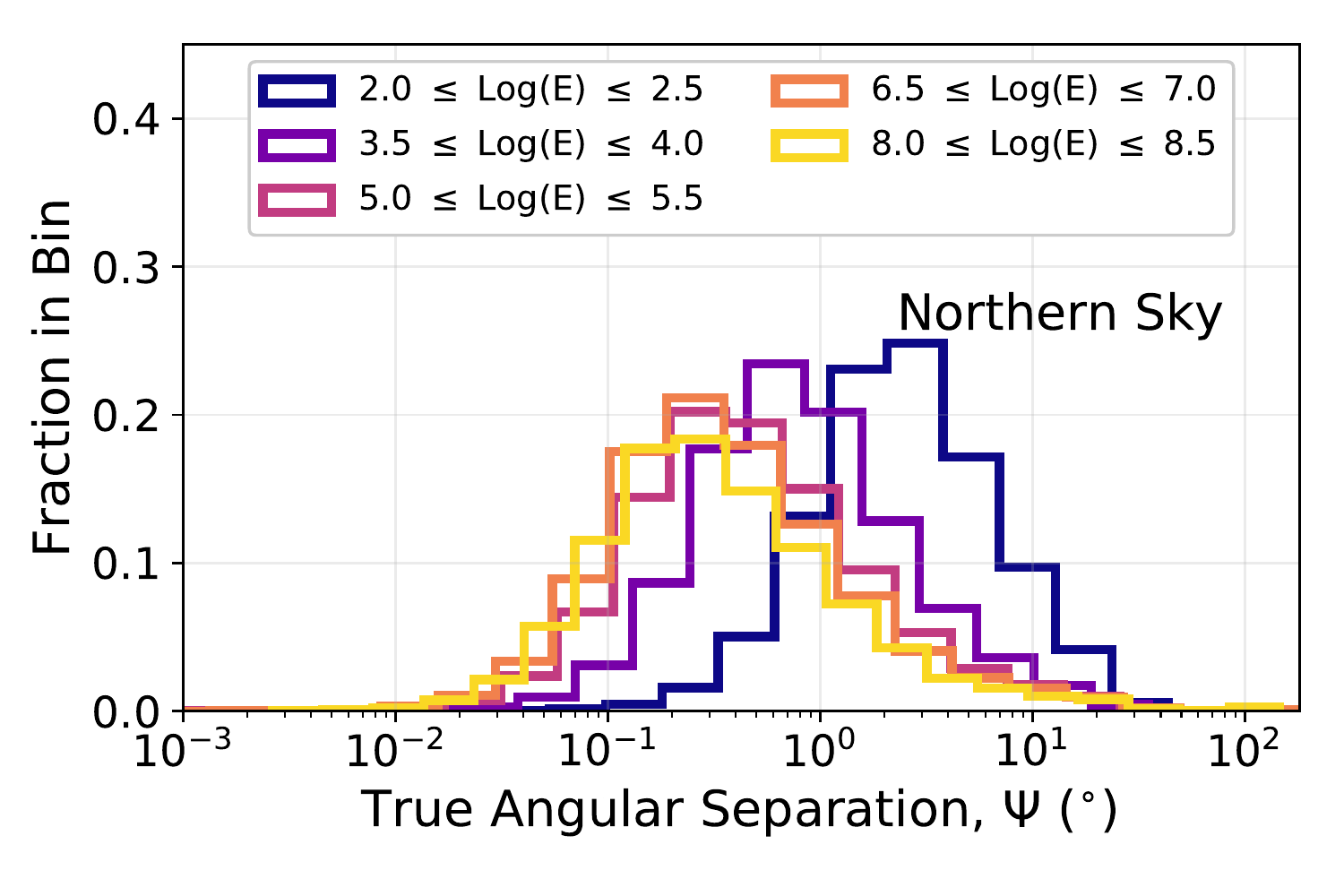}
\end{minipage}
\begin{minipage}[c][3.5cm][c]{0.32\textwidth}\centering
\includegraphics[width=\linewidth]{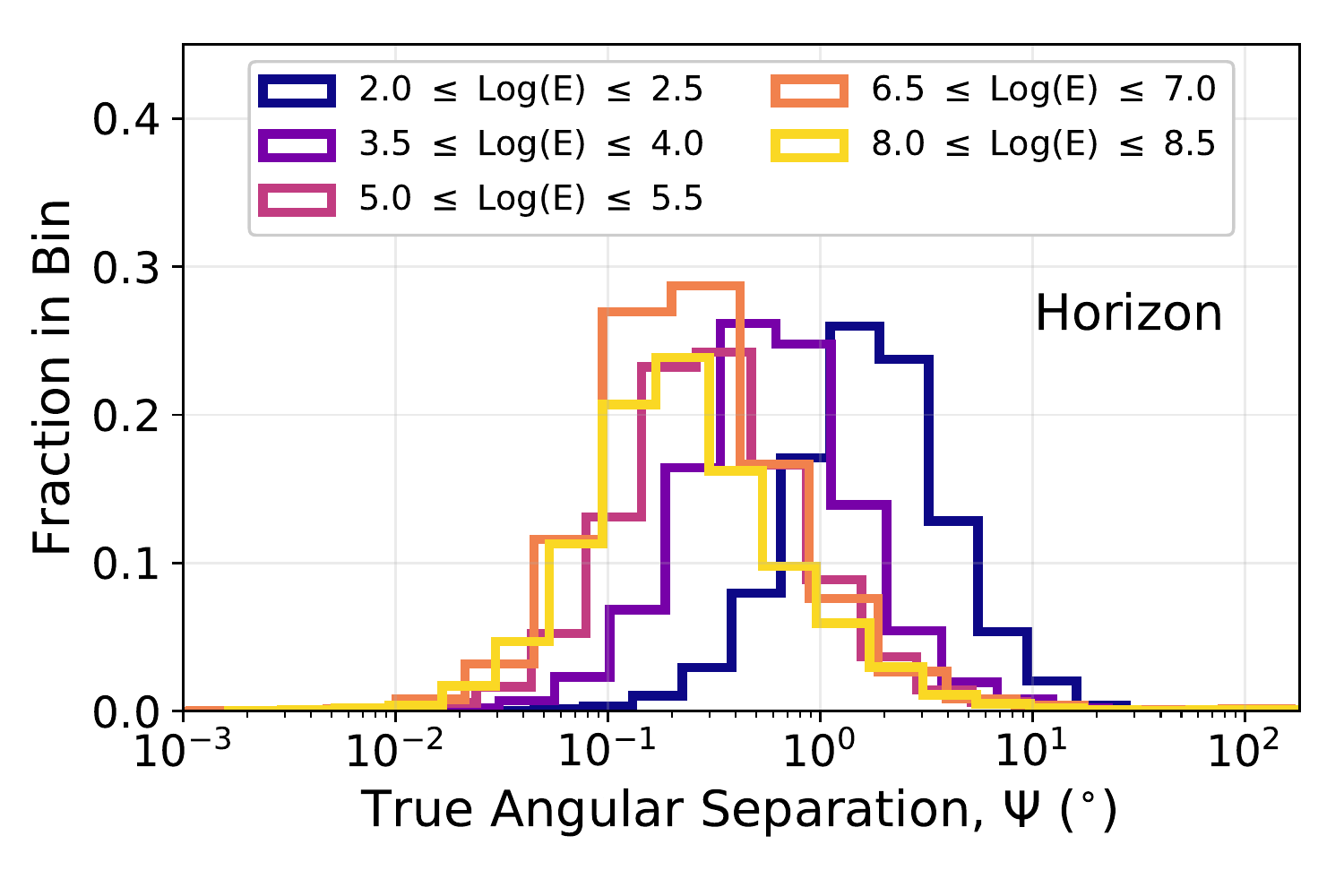}
\end{minipage}
\begin{minipage}[c][3.5cm][c]{0.32\textwidth}\centering
\includegraphics[width=\linewidth]{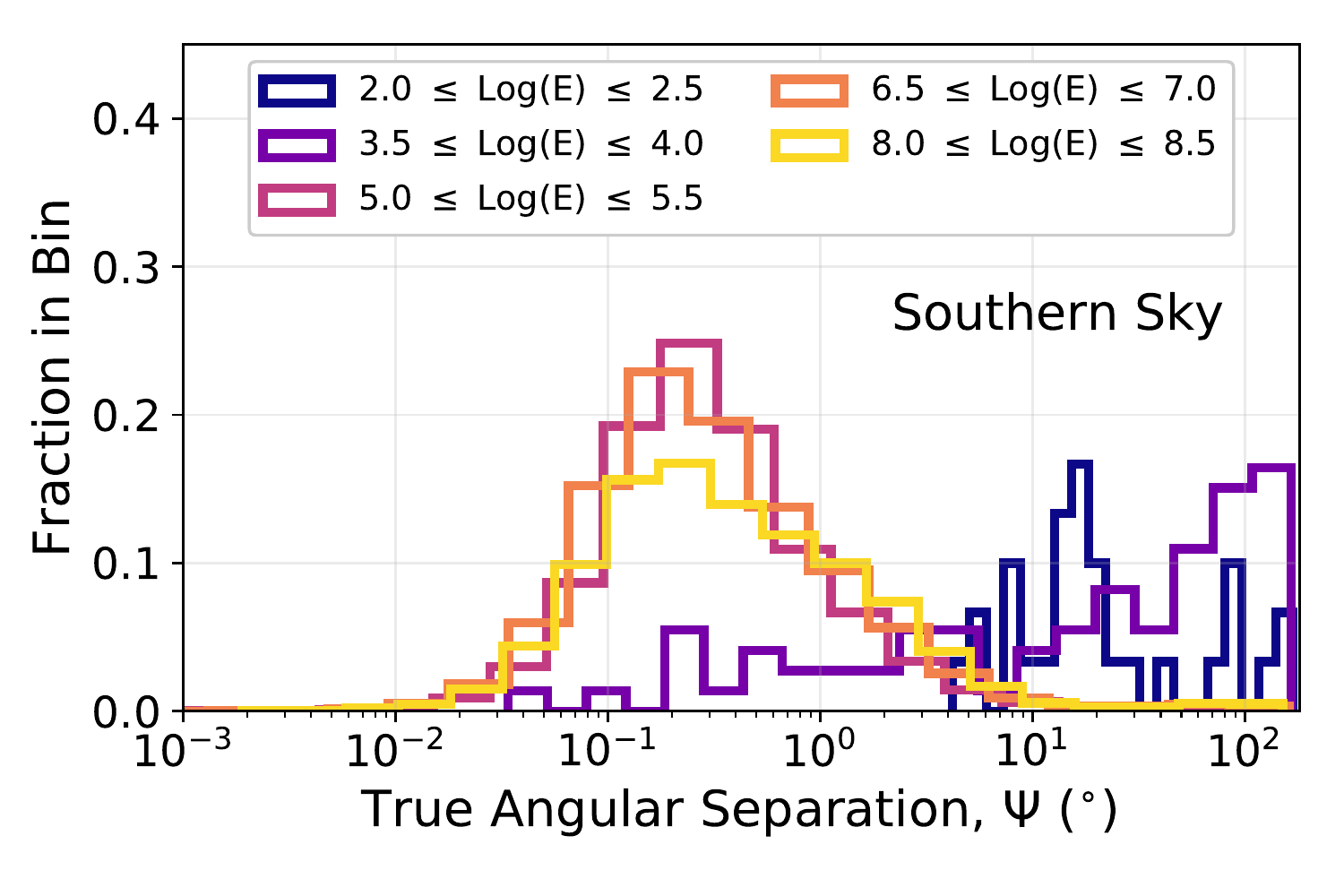}
\end{minipage}
\caption{The binned point spread functions measured for the Northern sky ($\delta<-10^\circ$), horizon ($-10^\circ\leq\delta<-10^\circ$), and Southern sky ($\delta\geq 10^\circ$) for IC86-II and later seasons. Each colored histogram corresponds to a different true neutrino energy range. For a falling $E^{-2}$ spectrum, most muons are reconstructed less than 1$^\circ$ from the neutrino origin.}\label{fig:PSF}
\end{figure*}

\begin{figure*}[t]\centering
\begin{minipage}[c][3.5cm][c]{0.32\textwidth}\centering
\includegraphics[width=\linewidth]{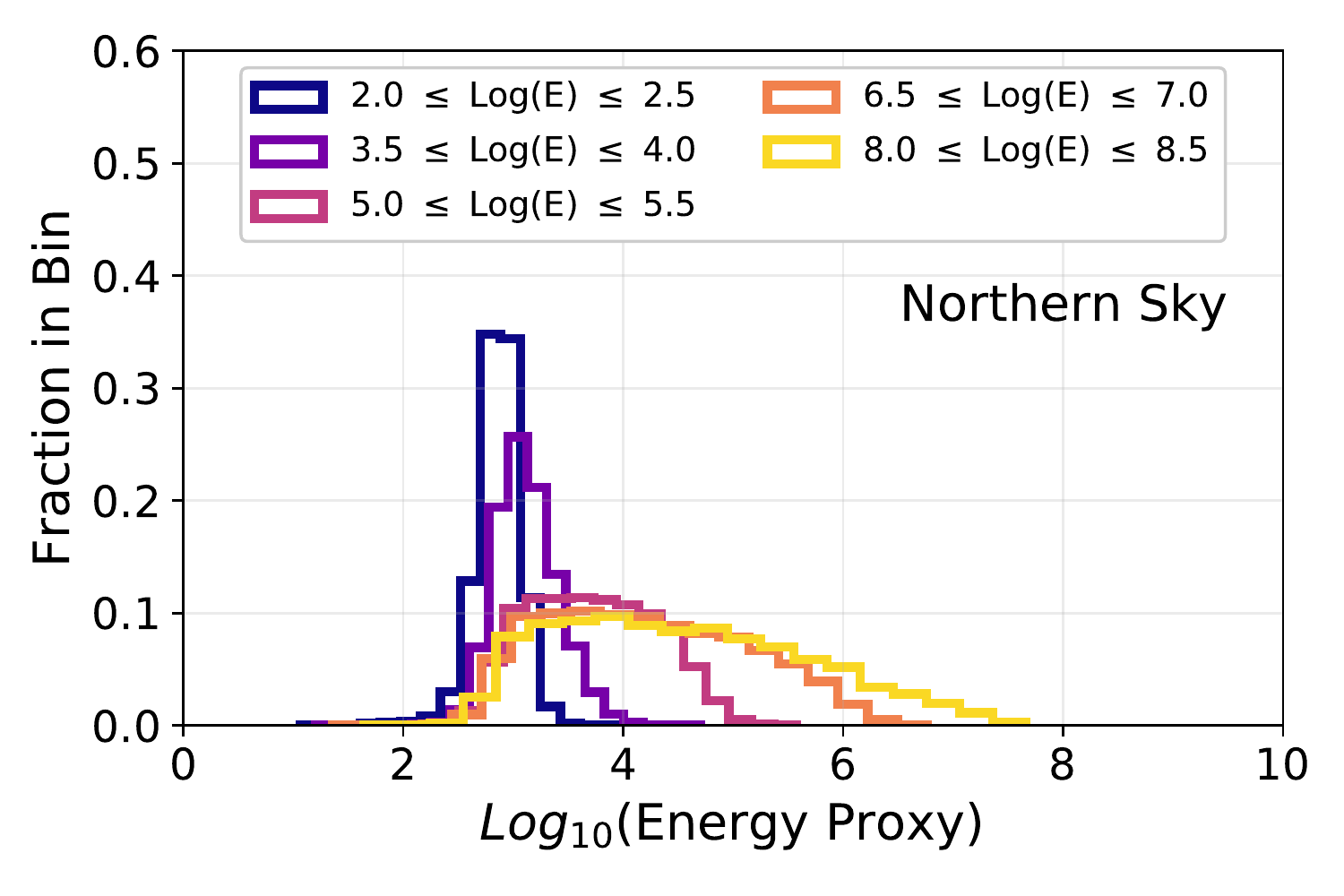}
\end{minipage}
\begin{minipage}[c][3.5cm][c]{0.32\textwidth}\centering
\includegraphics[width=\linewidth]{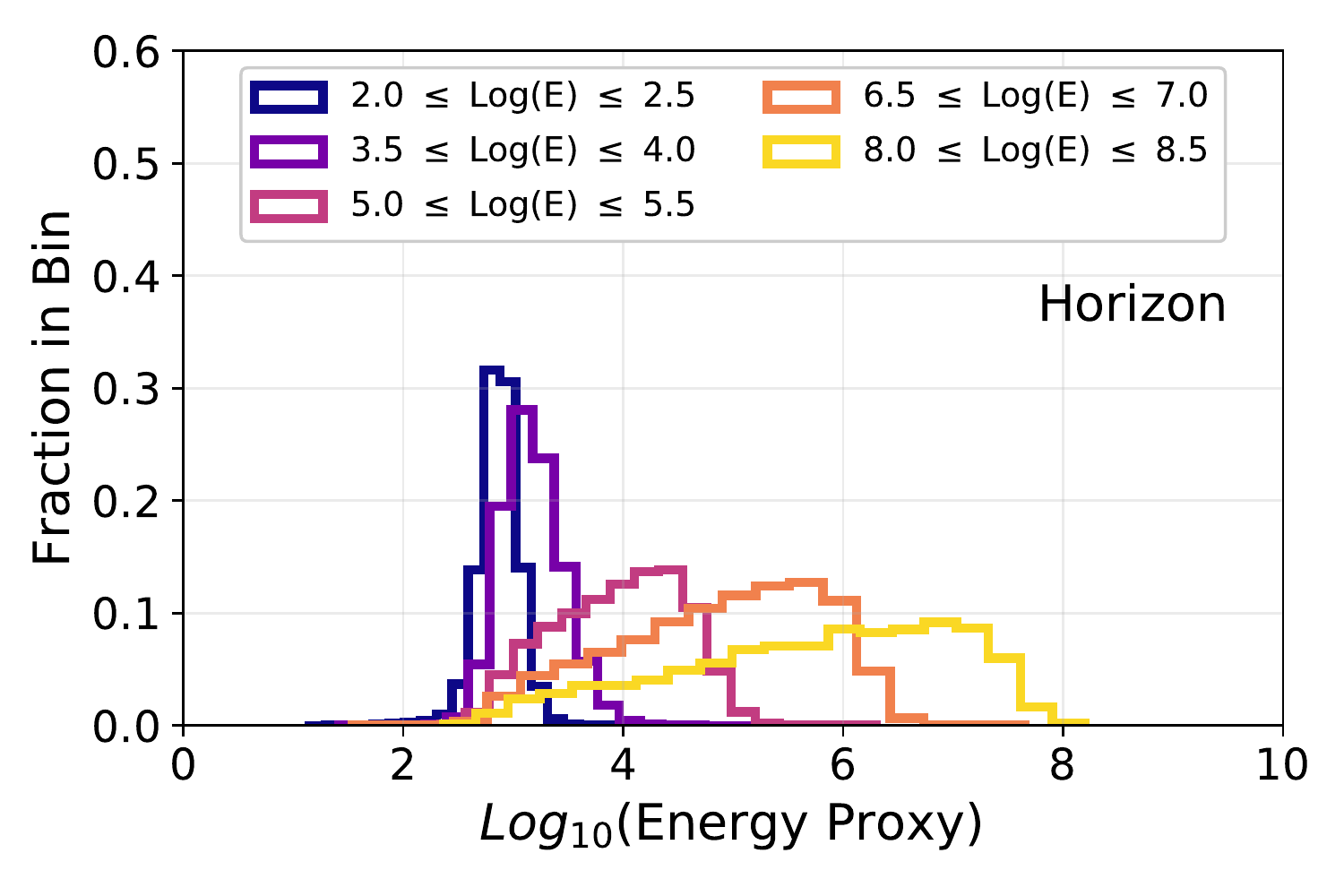}
\end{minipage}
\begin{minipage}[c][3.5cm][c]{0.32\textwidth}\centering
\includegraphics[width=\linewidth]{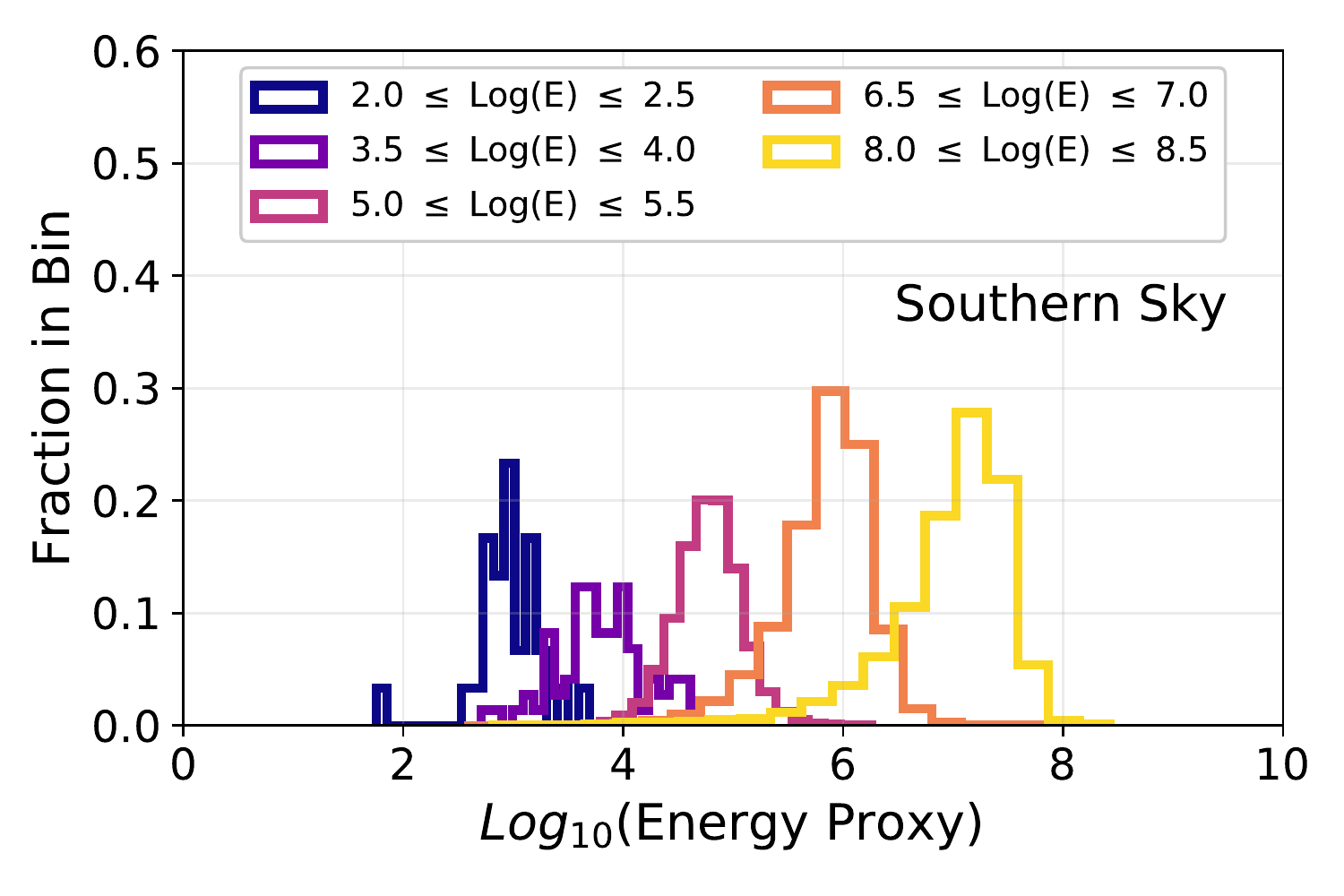}
\end{minipage}
\caption{The binned muon energy proxy reconstruction measured for the Northern sky ($\delta<-10^\circ$), horizon ($-10^\circ\leq\delta<-10^\circ$), and Southern sky ($\delta\geq 10^\circ$) for IC86-II and later seasons. Each colored histogram corresponds to a different true neutrino energy range. In the Southern sky, high energy events reconstruct near the incident neutrino energy. In the Northern sky and at the horizon, high energy events may interact far from the detector, producing energy losses which are not visible in IceCube. }\label{fig:ESmearing}
\end{figure*}

\begin{figure*}[t]\centering
\begin{minipage}[c][3.5cm][c]{0.32\textwidth}\centering
\includegraphics[width=\linewidth]{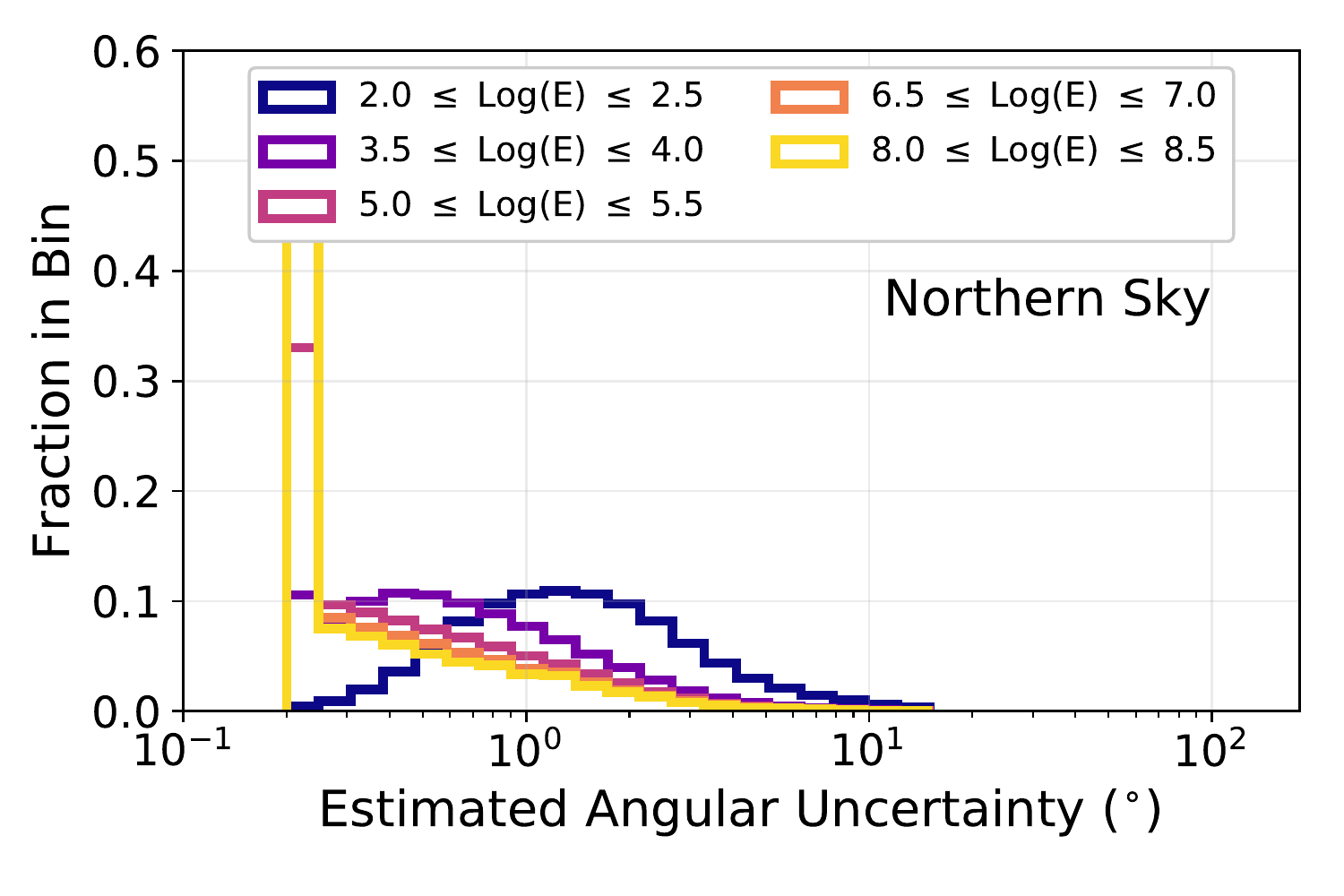}
\end{minipage}
\begin{minipage}[c][3.5cm][c]{0.32\textwidth}\centering
\includegraphics[width=\linewidth]{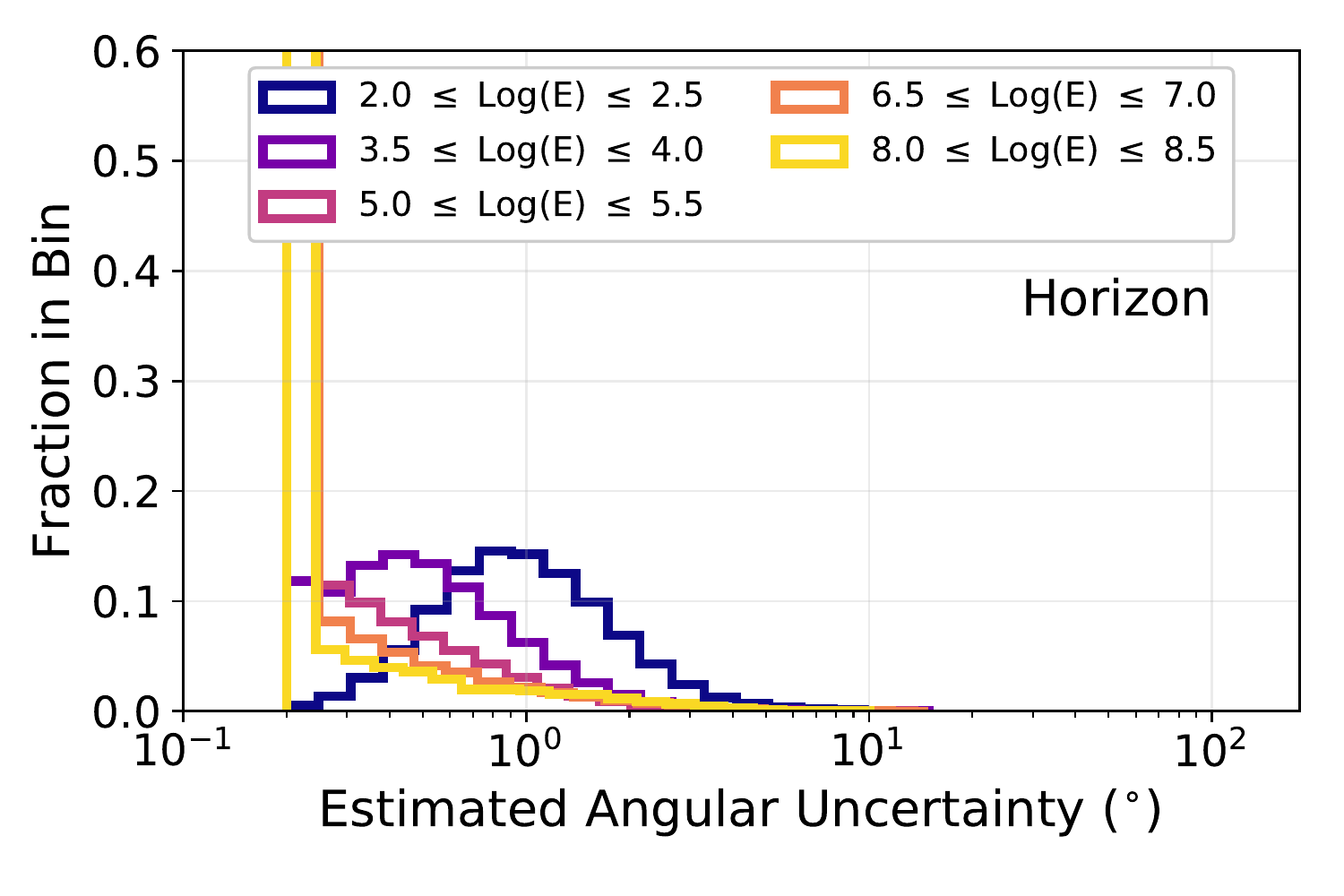}
\end{minipage}
\begin{minipage}[c][3.5cm][c]{0.32\textwidth}\centering
\includegraphics[width=\linewidth]{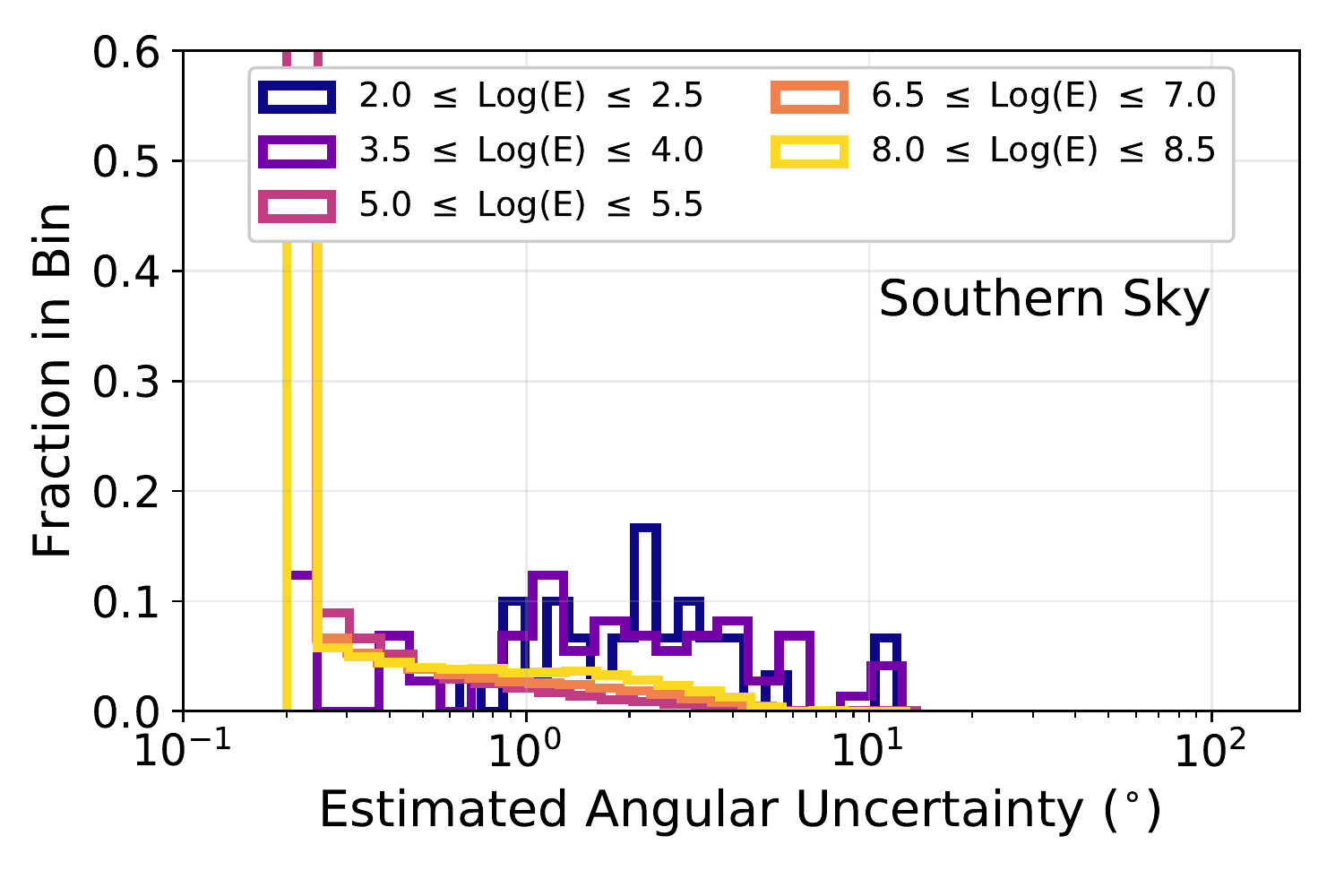}
\end{minipage}
\caption{The binned estimated angular uncertainty on the  reconstructed direction measured for the Northern sky ($\delta<-10^\circ$), horizon ($-10^\circ\leq\delta<-10^\circ$), and Southern sky ($\delta\geq 10^\circ$) for IC86-II and later seasons. Each colored histogram corresponds to a different true neutrino energy range. The estimated angular uncertainties have been calibrated to give correct coverage for an $E^{-2}$ spectrum. In order to avoid unaccounted-for uncertainties and to ensure no single event dominates our likelihood, a floor of 0.2$^\circ$ is included for all events.}\label{fig:AngErr}
\end{figure*}

Data seasons IC86-II through IC86-VII use multi-variate Boosted Decision Trees (BDT) to reduce the background of atmospheric muons and cascade events. Previous searches have shown the benefit of BDTs in the Northern sky \cite{Aartsen:2016oji,Aartsen:2018ywr}. In {\tt PSTracks v3}, a single BDT is trained for the Northern sky to recognize three classes of events: single muon tracks from atmospheric and astrophysical neutrinos, atmospheric muons, and cascades; neutrino-induced tracks are treated as signal. This BDT uses 11 variables related to event topology and reconstruction quality. The Northern BDT preserves $\sim90\%$ of the atmospheric neutrinos and $\sim0.1\%$ of the atmospheric muons from the initial selection of track-like events~\cite{Aartsen:2019fau}.

In the Southern hemisphere, BDTs are used to select only the best-reconstructed track-like events at the highest energies. In addition, the BDTs use four variables related to deposited energy along the track, as well as the light-arrival time of photons at the DOMs~\cite{Aartsen:2014cva,Aartsen:2016oji}. The large backgrounds of atmospheric muons and muon bundles require harsh cuts to reduce their rate significantly, resulting in an effective selection of only very high energy events. The selection effectively removes most Southern hemisphere events with an estimated energy below $\simeq10$~TeV; see Fig.~\ref{fig:enDist}. The IceTop surface array is used in addition as an active veto against coincident air-shower events for vertically down-going events~\cite{Aartsen:2013uuv}. 

The final all-sky event rate of about $4$~mHz is dominated by atmospheric muon neutrino interactions from the Northern hemisphere and by high-energy, well-reconstructed atmospheric muons in the Southern hemisphere. The preceding four years of data, collected with configurations IC40 through IC86-I, are handled exactly as in the past~\cite{Abbasi:2010rd,Aartsen:2013uuv,Schatto:2014kbj,Aartsen:2014cva}.


\section{Detector Response}\label{secIII}

Muon tracks induced by astrophysical neutrino interactions are the main signal category in the search for point-like sources of neutrinos. Detailed Monte Carlo simulation is used to evaluate the response of IceCube to such events and distinguish them from atmospheric backgrounds. These simulations may be characterized by a combination of the effective areas ($A_\mathrm{eff}$) and the reconstruction response functions. 

The number of expected events $N_\nu$ is given by 
\begin{align}\label{eq:ev_rate}
N_\nu=\int dt \int d\Omega\int_0^\infty dE\,A_\mathrm{eff}\left(E, \Omega\right) \phi_\nu\left(E_\nu,\Omega, t\right)
\end{align}
The incident neutrino flux $\phi_\nu$ can have an assumed form or be derived from simulation; see \cite{Aartsen:2016xlq}. The effective area for each season varies as a function of neutrino energy and declination as shown in Figure~\ref{fig:effA}. Tabulated effective areas for each season are included in this data release.

Reconstruction of events in {\tt PSTracks} proceeds in three steps, each incorporating effects from modeling of the Antarctic glacial ice medium. To begin, the direction of origin of the the candidate muon is reconstructed from the observed timing and charge in the detector following the algorithm described in Section~8.1 of Ref.~\cite{Ahrens:2004direc}. The angular distance between the reconstructed muon direction and the true neutrino direction is described by the point spread function (``PSF''). Binned examples of IceCube's PSF are shown in Figure~\ref{fig:PSF}.

The total energy loss of the muon track is then estimated following the description in section 9.1 of Ref.~\cite{Aartsen:2014erec}. The energy reconstruction yields a proxy for the muon energy at detector entry and a lower limit on the candidate neutrino energy. The observed distribution of the energy proxy can vary significantly for different declinations. For the Southern sky, observed muons from muon neutrino charged current interactions occur near the detector, giving an energy proxy close to the original neutrino's energy. For the Northern sky, neutrinos may interact while crossing the Earth before reaching IceCube, leading to unobservable energy losses, particularly at high energies. The energy proxy reconstruction is shown for IC86-II and later seasons in Figure~\ref{fig:ESmearing}.

The final stage is the estimation of the reconstruction angular uncertainty. The reconstruction likelihood space near the best-fit direction is mapped and fit with a paraboloid following Ref.~\cite{Neunhoeffer:2006}. The two dimensional width of the paraboloid fit is circularized by averaging the width across the two axes. In cases where the paraboloid method fails, {\tt PSTracks} instead relies on the Cramer-Rao method described in Ref.~\cite{Lunemann:2013oya}. In order to correct for the kinematic angle between neutrino and muon direction and to ensure correct coverage, a correction as a function of reconstructed energy is applied assuming an $E^{-2}$ flux such that the median estimated uncertainty as a function of energy gives 50\% containment for a two dimensional Gaussian distribution. While this correction can change with differing flux assumptions, the impact is small. A lower limit of 0.2$^\circ$ is applied to all events in order to avoid strong impacts from mismodeling of ice properties and to ensure that no single event dominates our likelihood calculations. The reconstruction angular uncertainties are shown in Figure~\ref{fig:AngErr}.

Reconstructed quantities are used to build probability density functions that the point source analysis uses in applying a maximum likelihood method. See Ref.~\cite{Aartsen:2019fau} for details of the likelihood construction, which exploits the spectral and spatial differences of astrophysical neutrinos and atmospheric backgrounds. 

\section{Comparison to Previous Releases}\label{secIV}

There have been multiple previous IceCube data releases presenting track-like events in the Northern and Southern hemisphere, primarily for the purpose of neutrino astronomy ~\cite{Abbasi:2010rd,Aartsen:2013uuv,IceCube:2018,IceCube:2019,IceCube:TXS2018}. The dataset detailed in this document is the latest iteration of a high statistics sample of track-like events. In this way, this dataset can be considered to be a successor of the previous data releases~\cite{IC40data,IC59data,IceCube:2018}, providing an updated description of the data from 2010--2012~\cite{IceCube:2018}, as well as adding six additional years of data in 2012--2018. Notably, this sample was used for the IceCube 10-year all-sky time-integrated analysis detailed in Ref.~\cite{Aartsen:2019fau}.

This sample includes several improvements in addition to the standardization of the IC86 data taking periods published in Ref.~\cite{IceCube:2018}. In the latest version of the data sample, referred to here as  {\tt PSTracks v3}, the event classifier and sample pre-cuts have been altered to better reject cascade-like events and accept track-like events. Additionally, the angular reconstruction has been updated, with more than a 10\% improvement in angular resolution for events greater than 10 TeV~\cite{Aartsen:2019fau}. The energy proxy remains unchanged between the two versions of the sample.

\begin{figure*}[p]
\centering
\includegraphics[width=0.5\linewidth,viewport= 0 0 530 390, clip=false]{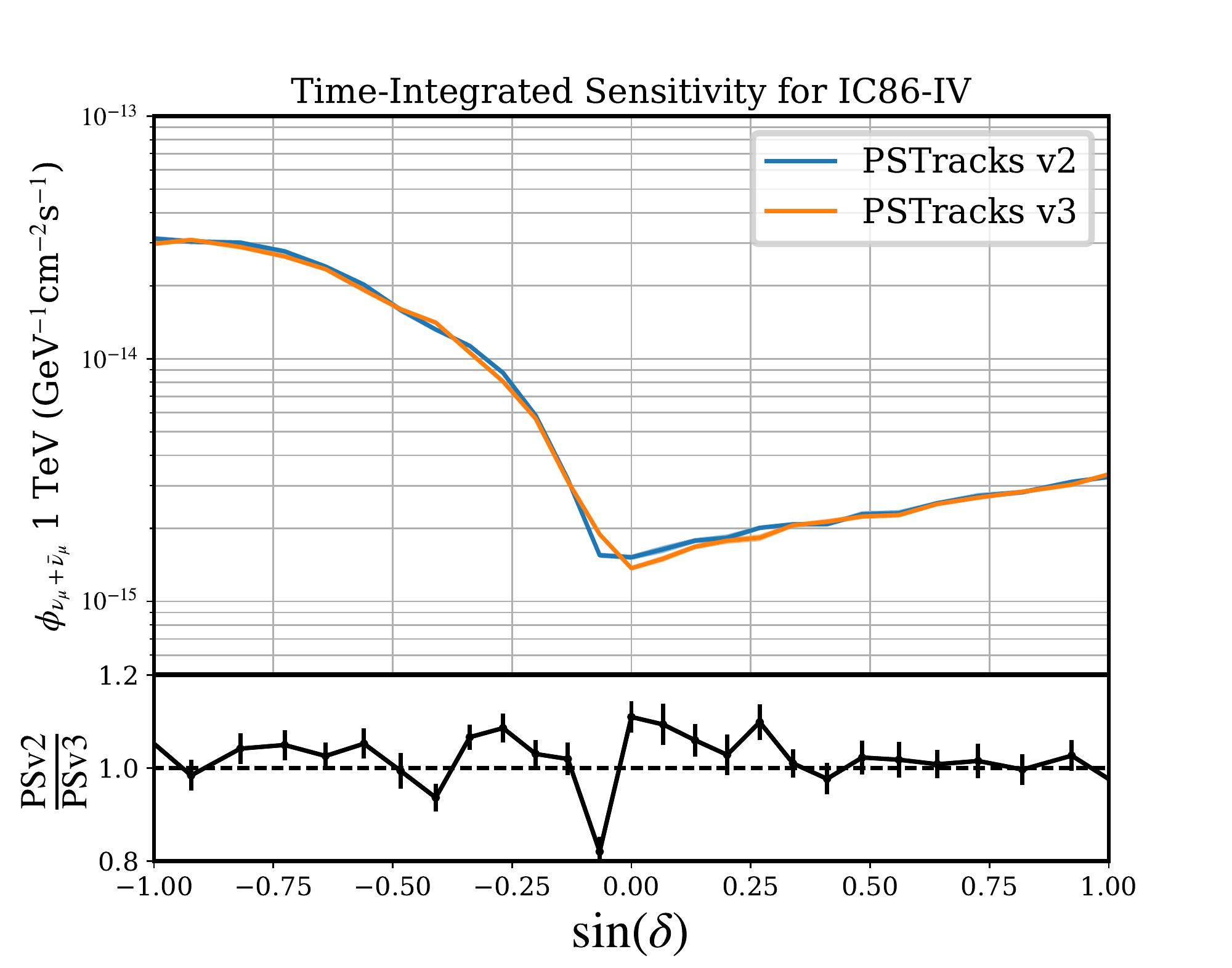}
\caption[]{A comparison of the 90\% C.L. sensitivity of a time-integrated point source search to an $E^{-2}$ source at various declinations. This is similar to the curves in figure 3 of Ref.~\cite{Aartsen:2019fau}, but here we calculate these values using only the IC86-IV season of both {\tt PSTracks v2} and {\tt PSTracks v3} for the purpose of comparing the two. For most declinations (particularly near the horizon), {\tt PSTracks v3} is a slight improvement over {\tt PSTracks v2}.}\label{fig:v2v3senscompare}
\end{figure*}

The net effect of the sample changes is an increase in event rate ($\sim$7\%) relative to previous versions of the sample. A year-by-year comparison of event counts between the older \MA{({\tt PSTracks v2})} and newer \MA{({\tt PSTracks v3})} versions of this sample can be seen in Table~\ref{tab:v2v3evtcontent}. A comparison of the one-year time-integrated sensitivity for the IC86-IV season for the two samples can be seen in Figure~\ref{fig:v2v3senscompare}. While the {\tt PSTracks v3} selection is expected to be, on average, more sensitive than the {\tt v2} selection, differences in event counts or reconstructions may lead to more varied changes in specific results. 

\begin{figure*}[htp]
\centering
\includegraphics[width=.45\linewidth]{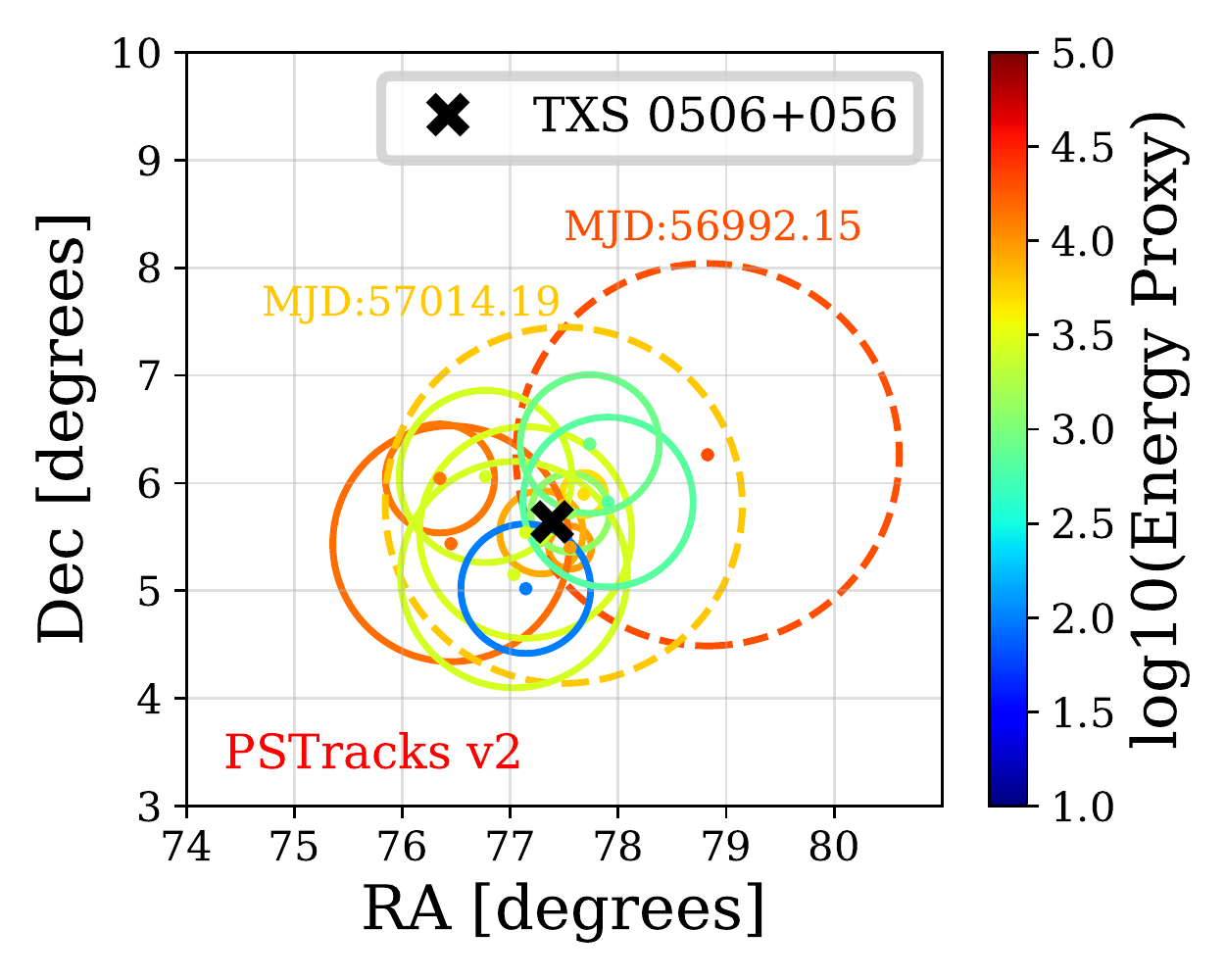}\hspace{0.2in}\includegraphics[width=.45\linewidth]{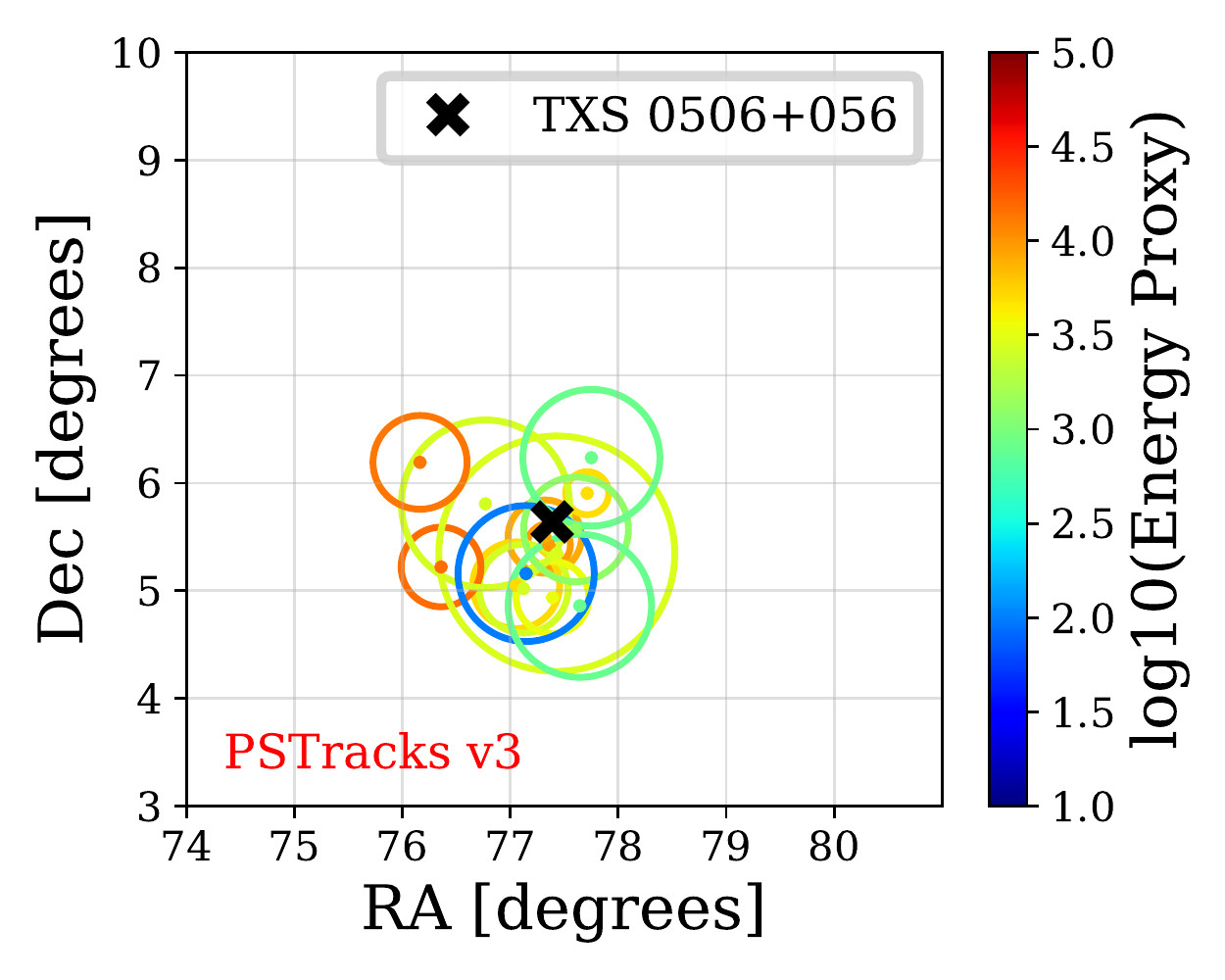}
\caption[]{The top 14 most signal-like flare events contributing to the 2014/2015 neutrino flare associated with TXS 0506+056 (black ``$\times$''), both in \MA{\tt PSTracks v2} (left, used for the analysis published in Ref.~\cite{IceCube:2018cha}), and the updated sample presented in this document ({\tt PSTracks v3}, right). The two cascade-like events present in {\tt v2} that were subsequently removed in {\tt v3} are shown as dashed circles in the plot on the left. The size of the colored circles corresponds to the $1\sigma$ containment region of the angular reconstruction of that event, and the color of the circle corresponds to the reconstructed energy proxy.}\label{fig:TXSEvtsCompare}
\end{figure*}

As an example of how the changing content of the different versions of this data sample can affect the process of searching for hotspots in IceCube data, we examine the 2014/2015 neutrino flare associated with TXS 0506+056~\cite{IceCube:2018cha}. The $3.5 \sigma$ excess seen in 2014/2015~\cite{IceCube:2018cha} was originally identified using the data sample corresponding to the data release Ref.~\cite{IceCube:2019}. Here, we repeat the untriggered flare search analysis, originally performed in Ref.~\cite{IceCube:2018cha}, but instead use the newest version of the data sample presented in this document, {\tt PSTracks v3}. The results of this cross-check can be seen in Table~\ref{tab:TXSCrossChecks}. Notably, the significance of the 2014/2015 neutrino flare decreases from $p=7.0 \times 10^{-5}$ ({\tt PSTracks v2}) to $p=8.1 \times 10^{-3}$ ({\tt PSTracks v3}) when using the most recent version of the data sample. A comparison of the reconstructed parameters of the most signal-like neutrino events contributing to the 2014/2015 flare in both versions of the data sample can be seen in Table~\ref{tab:TXSFlareEvtsTable} and Figure~\ref{fig:TXSEvtsCompare}.

The lower significance of the 2014/2015 neutrino flare in the most recent version of the sample has mainly been caused by the absence of two cascade-like events, occurring at MJD=56992.1586 and MJD=57014.1910, that existed in the {\tt PSTracks v2} data sample~\cite{IceCube:2019} associated with Ref.~\cite{IceCube:2018cha}, but have been removed from the {\tt PSTracks v3} data sample presented in this document. While cascade-like events may be used to search for point sources, they provide worse localization than track-like events and can provide additional backgrounds. In {\tt PSTracks v2}, a contribution from cascade-like events passing the selection was included in both the background and signal modeling. In PSTracks v3, events must pass a track length pre-cut requiring that all northern-sky events have a reconstructed track length greater than 200 meters, reducing the contribution from cascade events. As the two events at MJD=56992.1586  and  MJD=57014.1910 have reconstructed track lengths less than 200 meters, they are removed from {\tt PSTracks v3} prior to the application of the BDT. Performing the untriggered flare search using {\tt PSTracks v2} (Ref.~\cite{IceCube:2019}) with the two cascade events manually removed results in a drop in significance similar to that which is observed when using {\tt PSTracks v3}, as seen in Table~\ref{tab:TXSCrossChecks}. The drop in significance cannot be otherwise adequately explained by changes to the angular reconstruction or other differences between the two versions of the sample. 

It should be noted that the apparent drop in significance of the 2014/2015 neutrino flare is not entirely unprecedented. A reduction in the pre-trials significance of a time-integrated analysis preformed only at the location of TXS 056+056 is also seen in Ref.~\cite{Aartsen:2019fau} when using the newer version of this sample: $p=2.0 \times 10^{-5}$~($4.1\sigma$) with \MA{\tt PSTracks v2} ~\cite{IceCube:2018cha}, and $p=1.9 \times 10^{-4}$~($3.5\sigma$) with \MA{\tt PSTracks v3} ~\cite{Aartsen:2019fau}. Note that these significances do not include corrections for the number of searches (e.g. neither of these values are penalized for the fact that we have examined TXS 0506+056 multiple times, nor the fact that this same framework has been applied to other source locations), as in Ref.~\cite{Aartsen:2019fau} and Ref.~\cite{IceCube:2018cha}, and are simply used to compare the results of similar statistical frameworks applied to both {\tt PSTracks v2} and {\tt PSTracks v3}.
 
While both the time integrated and untriggered flare results associated with TXS 0506+056 appear to be less significant when using the newer version of this data sample, it is important to recognize that the results presented above are \textit{a posteriori} cross checks preformed in order to explore the effects of the altered content of {\tt PSTracks v3} in comparison to previous versions. Because {\tt PSTracks v2} and {\tt PStracks v3} differ in the specific event content and reconstructions, some fluctuations in significance of source candidates between versions is expected. Both samples are self-consistent, with {\tt PSTracks v3} expected to show better sensitivity to a generic, all-sky time-integrated analysis. For this reason, {\tt PSTracks v3} is preferred in the general use case.

\begin{figure*}[htp!]
\centering
\includegraphics[width=0.8\linewidth, clip=false]{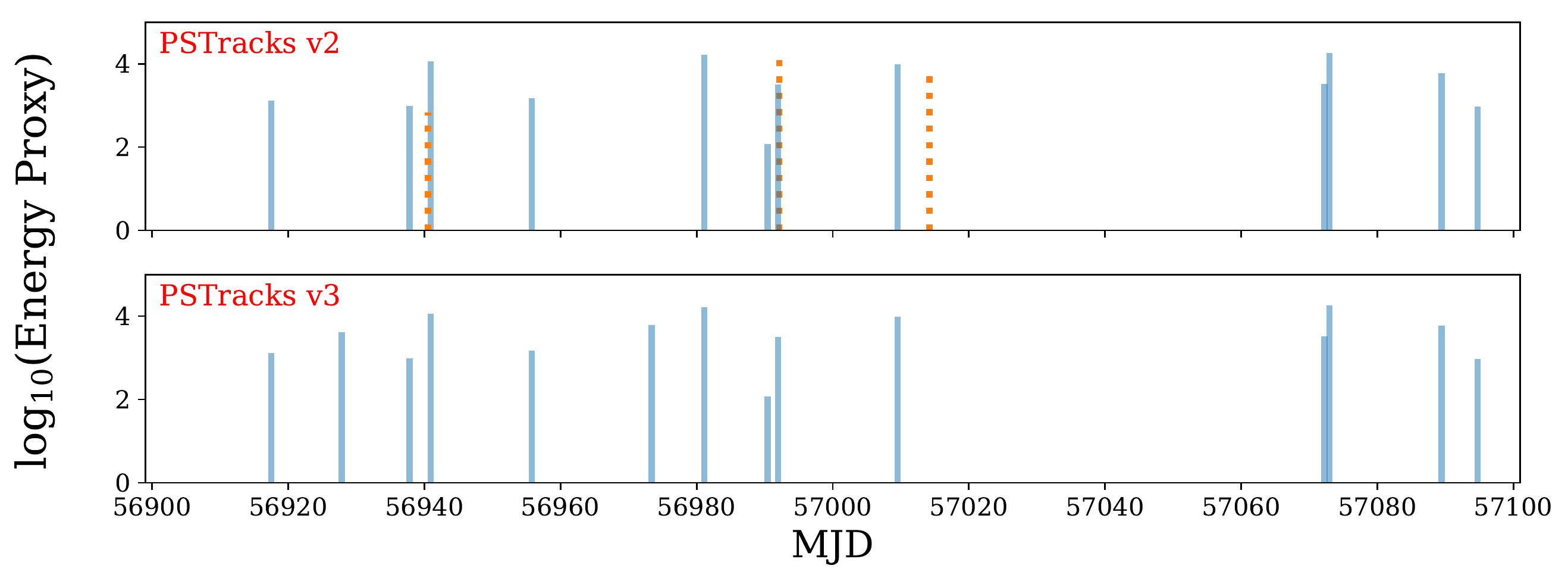}
\caption[]{The approximate time window of the 2014 neutrino flare associated with TXS 0506+056. Events listed in Table~\ref{tab:TXSFlareEvtsTable} are plotted as vertical lines with height equal to the event reconstructed energy proxy. Events present in {\tt PSTracks v2}, but not {\tt PSTracks v3} are shown as dotted orange lines.}\label{fig:txs_tevts}
\end{figure*}

\begin{table*}[p]
\centering
\begin{ruledtabular}
\begin{tabular}{lllccccc}
\multicolumn{8}{c}{Event Comparison Between \MA{\tt PSTracks~v2} and \MA{\tt PSTracks~v3}} \\[0.1cm]
Season & Start & End & Livetime & \MA{\tt PSTracks v2} & \MA{\tt PSTracks v3} & {\tt v2}$\rightarrow${\tt v3} Overlap & {\tt v3}$\rightarrow${\tt v2} Overlap \\ 
IC86-II & 2012/04/26\footnotemark & 2013/05/02 & 332.0d & 105300 & 112858 & 88.0\% & 82.1\% \\
IC86-III & 2013/05/02 & 2014/05/06 & 362.9d & 114834 & 122541 & 87.8\% & 82.3\% \\
IC86-IV & 2014/04/10\footnotemark & 2015/05/18 & 370.7d & 118456 & 127045 & 87.9\% & 82.0\% \\
\end{tabular}
\footnotetext[1]{Start date for test runs of the new processing. The remainder of this season began 2012/05/15}
\footnotetext[2]{Start date for test runs of the new processing. The remainder of this season began 2014/05/06}
\end{ruledtabular}
\caption[]{A comparison of the sample content between \MA{\tt PSTracks v2} and \MA{\tt PSTracks v3} for the period of overlap between the two samples. The rightmost two columns report the overlap in event content between the old and new versions of the samples for periods of shared livetime. \MA{The entry ``{\tt v2}$\rightarrow${\tt v3}''} refers to the percentage of events in {\tt PSTracks v2} that are also in \MA{\tt PSTracks v3}, and \MA{``{\tt v3}$\rightarrow${\tt v2}''} is the reverse. The \MA{older} version ({\tt v2}) of the sample has been discontinued as of the IC86-2015 season, having been replaced by \MA{\tt PSTracks v3}. Seasons prior to IC86-II are identical between the two versions of the sample, and seasons after IC86-IV only exist in the \MA{newer} ({\tt v3}) version of the sample.}\label{tab:v2v3evtcontent}
\end{table*}

\begin{table*}[p]
\centering
\begin{ruledtabular}
\begin{tabular}{cccccc}
\multicolumn{6}{c}{Untriggered Flare Cross-check Results} \\[0.1cm]
Sample & p-value (pre-trial) & $T_\text{start}$ & $T_\text{stop}$ & $n_s$ & $\gamma$ \\ 
\MA{\tt PSTracks v2}~\cite{IceCube:2018cha,IceCube:2019} & 7.0e-5 & 56937.81 & 57096.22 & 14.39 & 2.20  \\
\MA{\tt PSTracks v2} w/o cascades & 1.17e-3 & 56937.81 & 57112.65 & 12.22 & 2.26 \\
\MA{\tt PSTracks v3} (this release) & 8.14e-3 & 56927.86 & 57116.76 & 11.87 & 2.22\\
\end{tabular}
\end{ruledtabular}
\caption[]{The results of repeating the untriggered flare analysis preformed in ~\cite{IceCube:2018cha}, but using {\tt PSTracks v3} in place of {\tt PSTracks v2} ~\cite{IceCube:2019}, the dataset that was originally used. The apparent drop in significance when using {\tt PSTracks v3} can be explained by cascade-like events present in {\tt v2} that have been removed from {\tt v3}.}\label{tab:TXSCrossChecks}
\end{table*}

\begin{table*}[p]
\centering
\begin{ruledtabular}
\begin{tabular}{c|cccc|cccc}
\multicolumn{9}{c}{Events Contributing to the 2014 TXS 0506+056 Neutrino Flare} \\[0.1cm]
\hline
\multicolumn{1}{c|}{} &
\multicolumn{4}{c|}{\MA{\tt PSTracks v2}~\cite{IceCube:2018cha,IceCube:2019}} & 
\multicolumn{4}{c}{\MA{\tt PSTracks v3} (this release)}\\[0.1cm]
MJD & RA (deg) & Dec (deg) & $\sigma$ (deg) & $\log_{10}(E/{\rm GeV})$ & RA (deg) & Dec (deg) & $\sigma$ (deg) & $\log_{10}(E/{\rm GeV})$\\ 
\hline
56940.9084 & 77.55 & 5.40 & 0.20 & 3.97 & 77.35 & 5.42 & 0.20 & 3.97\\
57009.5301 & 77.28 & 5.54 & 0.38 & 3.91 & 77.32 & 5.50 & 0.34 & 3.91\\
57089.4395 & 77.68 & 5.89 & 0.20 & 3.69 & 77.71 & 5.90 & 0.20 & 3.69\\
57072.9895 & 76.45 & 5.43 & 1.09 & 4.17 & 76.35 & 5.22 & 0.36 & 4.17\\
56992.1586 & 78.82 & 6.26 & 1.77 & 4.30 & - & - & - & - \\
56981.1313& 76.34 & 6.04 & 0.50 & 4.13 & 76.16 & 6.19 & 0.43 & 4.13 \\
56955.7917 & 77.55 & 5.72 & 0.36 & 3.09 & 77.60 & 5.56 & 0.48 & 3.09 \\
57014.1910 & 77.49 & 5.79 & 1.65 & 3.79 & - & - & - & -  \\
57112.6530 & 77.14 & 5.54 & 0.98 & 3.46 & 77.43 & 5.34 & 1.09 & 3.46 \\
56991.9383 & 76.77 & 6.06 & 0.79 & 3.42 & 76.77 & 5.80 & 0.77 & 3.42 \\
57072.2089 & 77.03 & 5.14 & 1.05 & 3.43 & 76.35 & 5.22 & 0.36 & 3.43 \\
56990.4325 & 77.14 & 5.01 & 0.60 & 1.99 & 77.14 & 5.15 & 0.63 & 1.99 \\
56940.5215 & 77.90 & 5.82 & 0.78 & 2.82 & - & - & - & - \\
56937.8190 & 77.73 & 6.36 & 0.64 & 2.91 & 77.75 & 6.23 & 0.63 & 2.91 \\
56973.3971 & - & - & - & - & 77.05 & 5.05 & 0.40 & 3.71 \\
56927.8601 & - & - & - & - & 77.39 & 4.93 & 0.33 & 3.53 \\
56917.5296 & 78.27 & 5.86 & 0.54 & 3.05 & 78.32 & 5.85 & 0.52 & 3.05 \\
\end{tabular}
\end{ruledtabular}
\caption[]{The most signal-like events contributing to the 2014/2015 TXS 0506+056 neutrino flare in, both, the data sample published in Ref.~\cite{IceCube:2019} ({\tt PSTracks v2}) as well as the data sample presented here ({\tt PSTracks v3}). There were no changes to energy reconstruction between the two sample versions, but the angular reconstruction and sample event content have changed. The events occurring at MJD=56992.1586 and MJD=57014.1910 are cascade events that have been treated as tracks by the reconstruction, and as such the direction and energy information reported in the {\tt PSTracks v2} sample is unlikely to be a good description of the true event properties.}\label{tab:TXSFlareEvtsTable}
\end{table*}

\begin{acknowledgements}
\input{agencies.tex}
\end{acknowledgements}

\bibliographystyle{utphys_mod}

\bibliography{references}

\end{document}

%% file: authorlist.tex
\affiliation{III. Physikalisches Institut, RWTH Aachen University, D-52056 Aachen, Germany}
\affiliation{Department of Physics, University of Adelaide, Adelaide, 5005, Australia}
\affiliation{Dept. of Physics and Astronomy, University of Alaska Anchorage, 3211 Providence Dr., Anchorage, AK 99508, USA}
\affiliation{Dept. of Physics, University of Texas at Arlington, 502 Yates St., Science Hall Rm 108, Box 19059, Arlington, TX 76019, USA}
\affiliation{CTSPS, Clark-Atlanta University, Atlanta, GA 30314, USA}
\affiliation{School of Physics and Center for Relativistic Astrophysics, Georgia Institute of Technology, Atlanta, GA 30332, USA}
\affiliation{Dept. of Physics, Southern University, Baton Rouge, LA 70813, USA}
\affiliation{Dept. of Physics, University of California, Berkeley, CA 94720, USA}
\affiliation{Lawrence Berkeley National Laboratory, Berkeley, CA 94720, USA}
\affiliation{Institut f{\"u}r Physik, Humboldt-Universit{\"a}t zu Berlin, D-12489 Berlin, Germany}
\affiliation{Fakult{\"a}t f{\"u}r Physik {\&} Astronomie, Ruhr-Universit{\"a}t Bochum, D-44780 Bochum, Germany}
\affiliation{Universit{\'e} Libre de Bruxelles, Science Faculty CP230, B-1050 Brussels, Belgium}
\affiliation{Vrije Universiteit Brussel (VUB), Dienst ELEM, B-1050 Brussels, Belgium}
\affiliation{Department of Physics and Laboratory for Particle Physics and Cosmology, Harvard University, Cambridge, MA 02138, USA}
\affiliation{Dept. of Physics, Massachusetts Institute of Technology, Cambridge, MA 02139, USA}
\affiliation{Dept. of Physics and Institute for Global Prominent Research, Chiba University, Chiba 263-8522, Japan}
\affiliation{Department of Physics, Loyola University Chicago, Chicago, IL 60660, USA}
\affiliation{Dept. of Physics and Astronomy, University of Canterbury, Private Bag 4800, Christchurch, New Zealand}
\affiliation{Dept. of Physics, University of Maryland, College Park, MD 20742, USA}
\affiliation{Dept. of Astronomy, Ohio State University, Columbus, OH 43210, USA}
\affiliation{Dept. of Physics and Center for Cosmology and Astro-Particle Physics, Ohio State University, Columbus, OH 43210, USA}
\affiliation{Niels Bohr Institute, University of Copenhagen, DK-2100 Copenhagen, Denmark}
\affiliation{Dept. of Physics, TU Dortmund University, D-44221 Dortmund, Germany}
\affiliation{Dept. of Physics and Astronomy, Michigan State University, East Lansing, MI 48824, USA}
\affiliation{Dept. of Physics, University of Alberta, Edmonton, Alberta, Canada T6G 2E1}
\affiliation{Erlangen Centre for Astroparticle Physics, Friedrich-Alexander-Universit{\"a}t Erlangen-N{\"u}rnberg, D-91058 Erlangen, Germany}
\affiliation{Physik-department, Technische Universit{\"a}t M{\"u}nchen, D-85748 Garching, Germany}
\affiliation{D{\'e}partement de physique nucl{\'e}aire et corpusculaire, Universit{\'e} de Gen{\`e}ve, CH-1211 Gen{\`e}ve, Switzerland}
\affiliation{Dept. of Physics and Astronomy, University of Gent, B-9000 Gent, Belgium}
\affiliation{Dept. of Physics and Astronomy, University of California, Irvine, CA 92697, USA}
\affiliation{Karlsruhe Institute of Technology, Institut f{\"u}r Kernphysik, D-76021 Karlsruhe, Germany}
\affiliation{Dept. of Physics and Astronomy, University of Kansas, Lawrence, KS 66045, USA}
\affiliation{SNOLAB, 1039 Regional Road 24, Creighton Mine 9, Lively, ON, Canada P3Y 1N2}
\affiliation{Department of Physics and Astronomy, UCLA, Los Angeles, CA 90095, USA}
\affiliation{Department of Physics, Mercer University, Macon, GA 31207-0001, USA}
\affiliation{Dept. of Astronomy, University of Wisconsin{\textendash}Madison, Madison, WI 53706, USA}
\affiliation{Dept. of Physics and Wisconsin IceCube Particle Astrophysics Center, University of Wisconsin{\textendash}Madison, Madison, WI 53706, USA}
\affiliation{Institute of Physics, University of Mainz, Staudinger Weg 7, D-55099 Mainz, Germany}
\affiliation{Department of Physics, Marquette University, Milwaukee, WI, 53201, USA}
\affiliation{Institut f{\"u}r Kernphysik, Westf{\"a}lische Wilhelms-Universit{\"a}t M{\"u}nster, D-48149 M{\"u}nster, Germany}
\affiliation{Bartol Research Institute and Dept. of Physics and Astronomy, University of Delaware, Newark, DE 19716, USA}
\affiliation{Dept. of Physics, Yale University, New Haven, CT 06520, USA}
\affiliation{Dept. of Physics, University of Oxford, Parks Road, Oxford OX1 3PU, UK}
\affiliation{Dept. of Physics, Drexel University, 3141 Chestnut Street, Philadelphia, PA 19104, USA}
\affiliation{Physics Department, South Dakota School of Mines and Technology, Rapid City, SD 57701, USA}
\affiliation{Dept. of Physics, University of Wisconsin, River Falls, WI 54022, USA}
\affiliation{Dept. of Physics and Astronomy, University of Rochester, Rochester, NY 14627, USA}
\affiliation{Oskar Klein Centre and Dept. of Physics, Stockholm University, SE-10691 Stockholm, Sweden}
\affiliation{Dept. of Physics and Astronomy, Stony Brook University, Stony Brook, NY 11794-3800, USA}
\affiliation{Dept. of Physics, Sungkyunkwan University, Suwon 16419, Korea}
\affiliation{Institute of Basic Science, Sungkyunkwan University, Suwon 16419, Korea}
\affiliation{Dept. of Physics and Astronomy, University of Alabama, Tuscaloosa, AL 35487, USA}
\affiliation{Dept. of Astronomy and Astrophysics, Pennsylvania State University, University Park, PA 16802, USA}
\affiliation{Dept. of Physics, Pennsylvania State University, University Park, PA 16802, USA}
\affiliation{Dept. of Physics and Astronomy, Uppsala University, Box 516, S-75120 Uppsala, Sweden}
\affiliation{Dept. of Physics, University of Wuppertal, D-42119 Wuppertal, Germany}
\affiliation{DESY, D-15738 Zeuthen, Germany}

\author{R. Abbasi}
\affiliation{Department of Physics, Loyola University Chicago, Chicago, IL 60660, USA}
\author{M. Ackermann}
\affiliation{DESY, D-15738 Zeuthen, Germany}
\author{J. Adams}
\affiliation{Dept. of Physics and Astronomy, University of Canterbury, Private Bag 4800, Christchurch, New Zealand}
\author{J. A. Aguilar}
\affiliation{Universit{\'e} Libre de Bruxelles, Science Faculty CP230, B-1050 Brussels, Belgium}
\author{M. Ahlers}
\affiliation{Niels Bohr Institute, University of Copenhagen, DK-2100 Copenhagen, Denmark}
\author{M. Ahrens}
\affiliation{Oskar Klein Centre and Dept. of Physics, Stockholm University, SE-10691 Stockholm, Sweden}
\author{C. Alispach}
\affiliation{D{\'e}partement de physique nucl{\'e}aire et corpusculaire, Universit{\'e} de Gen{\`e}ve, CH-1211 Gen{\`e}ve, Switzerland}
\author{N. M. Amin}
\affiliation{Bartol Research Institute and Dept. of Physics and Astronomy, University of Delaware, Newark, DE 19716, USA}
\author{K. Andeen}
\affiliation{Department of Physics, Marquette University, Milwaukee, WI, 53201, USA}
\author{T. Anderson}
\affiliation{Dept. of Physics, Pennsylvania State University, University Park, PA 16802, USA}
% ...

%% file: agencies.tex
USA {\textendash} U.S. National Science Foundation-Office of Polar Programs,
U.S. National Science Foundation-Physics Division,
Wisconsin Alumni Research Foundation,
Center for High Throughput Computing (CHTC) at the University of Wisconsin{\textendash}Madison,
Open Science Grid (OSG),
Extreme Science and Engineering Discovery Environment (XSEDE),
U.S. Department of Energy-National Energy Research Scientific Computing Center,
Particle astrophysics research computing center at the University of Maryland,
Institute for Cyber-Enabled Research at Michigan State University,
and Astroparticle physics computational facility at Marquette University;
Belgium {\textendash} Funds for Scientific Research (FRS-FNRS and FWO),
FWO Odysseus and Big Science programmes,
and Belgian Federal Science Policy Office (Belspo);
Germany {\textendash} Bundesministerium f{\"u}r Bildung und Forschung (BMBF),
Deutsche Forschungsgemeinschaft (DFG),
Helmholtz Alliance for Astroparticle Physics (HAP),
Initiative and Networking Fund of the Helmholtz Association,
Deutsches Elektronen Synchrotron (DESY),
and High Performance Computing cluster of the RWTH Aachen;
Sweden {\textendash} Swedish Research Council,
Swedish Polar Research Secretariat,
Swedish National Infrastructure for Computing (SNIC),
and Knut and Alice Wallenberg Foundation;
Australia {\textendash} Australian Research Council;
Canada {\textendash} Natural Sciences and Engineering Research Council of Canada,
Calcul Qu{\'e}bec, Compute Ontario, Canada Foundation for Innovation, WestGrid, and Compute Canada;
Denmark {\textendash} Villum Fonden, Danish National Research Foundation (DNRF), Carlsberg Foundation;
New Zealand {\textendash} Marsden Fund;
Japan {\textendash} Japan Society for Promotion of Science (JSPS)
and Institute for Global Prominent Research (IGPR) of Chiba University;
Korea {\textendash} National Research Foundation of Korea (NRF);
Switzerland {\textendash} Swiss National Science Foundation (SNSF);
United Kingdom {\textendash} Department of Physics, University of Oxford.